\documentclass[review,12pt]{elsarticle}

\usepackage{setspace}
\usepackage{subcaption}
\usepackage{multirow}
\usepackage{multicol}
\usepackage{graphicx}
\RequirePackage{amsmath}
\usepackage{bm}
\usepackage{xfrac}
\usepackage{float}
\usepackage{arydshln} 

\usepackage{amsmath,esint,amsfonts}

\usepackage[most]{tcolorbox}
\usepackage[colorinlistoftodos]{todonotes}

\usepackage{lineno}
\usepackage[colorlinks=true,linkcolor=blue,urlcolor=blue,citecolor=blue,anchorcolor=blue]{hyperref}
\usepackage{cleveref}

\journal{Physics of Fluids}

\biboptions{sort&compress} 

\begin{document}

\begin{frontmatter}
	\title{Shear-driven flow in an elliptical enclosure generated by an inner rotating circular cylinder}

	\author{Akash Unnikrishnan$^{a,}$\fnref{Corresponding Author}}
	\author{Shantanu Shahane$^{b}$}
	\author{Vinod Narayanan$^{a}$}
	\author{Surya Pratap Vanka$^{c}$}
	\address{a Department of Mechanical Engineering,\\
		Indian Institute of Technology Gandhinagar,		Gandhinagar, Gujarat 382355, India}
	\address{b National Center for Supercomputing Applications,\\
        University of Illinois at Urbana-Champaign, Urbana, Illinois 61801, USA}
	\address{c Department of Mechanical Science and Engineering,\\
		University of Illinois at Urbana-Champaign, Urbana, Illinois 61801, USA}
	\fntext[Author to whom correspondence should be addressed]{\vspace{0.3cm}Corresponding Author Email Address: \url{akash.unnikrishnan@iitgn.ac.in}}

\begin{abstract}
    Shear-driven flow between a rotating cylinder and a stationary elliptical enclosure is studied in this paper. Two-dimensional time-dependent Navier Stokes equations are solved using a meshless method where interpolations are done with Polyharmonic Spline Radial Basis Functions. The fluid flow is analyzed for various aspect ratios of the ellipse and eccentric placements of the inner cylinder. Contour plots of vorticity with streamlines, plots of non-dimensional torque, and the angle of eye of the primary vortex are presented in the paper for Reynolds numbers between 200 and 2000. Formation of Moffatt like vortices in the wide-gap region of the model is observed and some benchmark data are provided for various cases that are simulated.
\end{abstract}


\end{frontmatter}

\section{INTRODUCTION}
Shear-driven flows in enclosures have been studied extensively both experimentally and computationally. The most widely studied geometry is a square cavity \cite{kuhlmann2019lid, shankar2000fluid}, in which the motion of the top wall results in multiple vortices including a large central vortex and multiple corner vortices at the bottom wall. The driven cavity flow has been widely computed by many researchers to validate new numerical algorithms and computer programs to solve the Navier-Stokes equations and compare results with benchmark solutions \cite{ghia1982high, botella1998benchmark}. The driven cavity problem has also been studied in three dimensions \cite{albensoeder2005accurate}, where Taylor-Gortler vortices form on the bottom wall parallel to the top moving boundary. As a variation to the square cavity, rectangular cavities of different aspect ratios have also been studied by \citet{cheng2006vortex}. In tall rectangular cavities driven by the shear of the top boundary, a series of vortices are formed in the vertical direction with their strength decreasing with depth from the top to the bottom boundary, ultimately yielding a stagnant region in the bottom.

Shear-driven flows in shaped cavities have also been studied numerically. \citet{darr1991separated} analyzed the formation of vortices in a trapezoidal cavity with top and bottom walls moving at different velocities and in the same or opposite directions. The Reynolds number based on the depth of the cavity is varied, resulting in complex formations of vortices. An efficient multigrid procedure on curvilinear grids is used to rapidly converge the equations on fine grids. \citet{mcquain1994steady} solved a fourth-order stream function equation to determine the velocities in a driven trapezoidal cavity. They also considered square and triangular cavities and presented results for different Reynolds numbers. \citet{jyotsna1995multigrid} used a Control Volume Finite Element Method (CVFEM) and a nested multigrid procedure to obtain benchmark solutions to shear driven flow in a triangular cavity. A series of Moffatt vortices \cite{Moffatt1964viscous} are captured in the triangular cavity at different Reynolds numbers with their strength progressively decaying toward the corner.  Driven cavities of triangular shape have also been studied by other researchers using different numerical methods [e.g. \cite{erturk2007fine, an2019lid, jagannathan2014spectral}]. \citet{vanka2008immersed} further studied shear-driven vortices in a number of complex enclosures with the flow driven by different side boundaries.  They used a Cartesian grid with the Immersed Boundary Method (IBM) at the boundaries. Very complex and interesting vortical flows are observed including a series of Moffatt vortices. Several other geometries including an arcsin cavity, an isosceles triangle cavity \cite{an2019lid}, a semi-elliptical cavity \cite{ren_guo_2017} and a semi-circular cavity have also been computed \cite{glowinski2006wall, mercan2009vortex, feng2019multi, scott2013Moffatt}. \citet{ozalp2010experimental} conducted measurements of flows in cavities of different shapes in a water tunnel with flow parallel to the flat surface. They considered triangular, rectangular and semi-circular cavities at different Reynolds numbers. The Particle Image Velocimetry (PIV) technique is used to map the streamlines and vorticity distributions. Driven cavity flows with power-law fluids, and magnetic fields have been reported in \cite{jin2017gpu, jin2015three}. A number of other studies of driven cavity flow, especially dealing with stability of the two-dimensional flow are reviewed by\citet{kuhlmann2019lid} and \citet{shankar2000fluid}.

Shear-driven flows in enclosures with moving inner boundaries also give rise to complex vortical structures \cite{paul2020conjugate} which are relatively unexplored. The commonly studied flow is that between two rotating cylinders for which axial vortical cells are formed. These cells known as Taylor-Couette cells arise out of three-dimensional instabilities on a base axisymmetric flow. However, when the inner cylinder is eccentrically placed, non-axisymmetric flow distributions with multiple vortices are observed \cite{koschmieder1976taylor,san1984flow,diprima1972flow}. Flow in an elliptical enclosure with a rotating inner cylinder has relevance to elliptical bearings, and several studies have computed the drag and lift in such flows \cite{HASHIMOTO1984,jose2018review}. They also have applications in electrochemical reactors \cite{gabe1998rotating}, rotary fractionating columns \cite{macleod1959performance}, polymer processing with screw extruders \cite{potente1994design}, etc. Understanding the flow structures is important to the design of these applications.

\begin{figure}[h]
    \centering
    \includegraphics[width=0.7\textwidth]{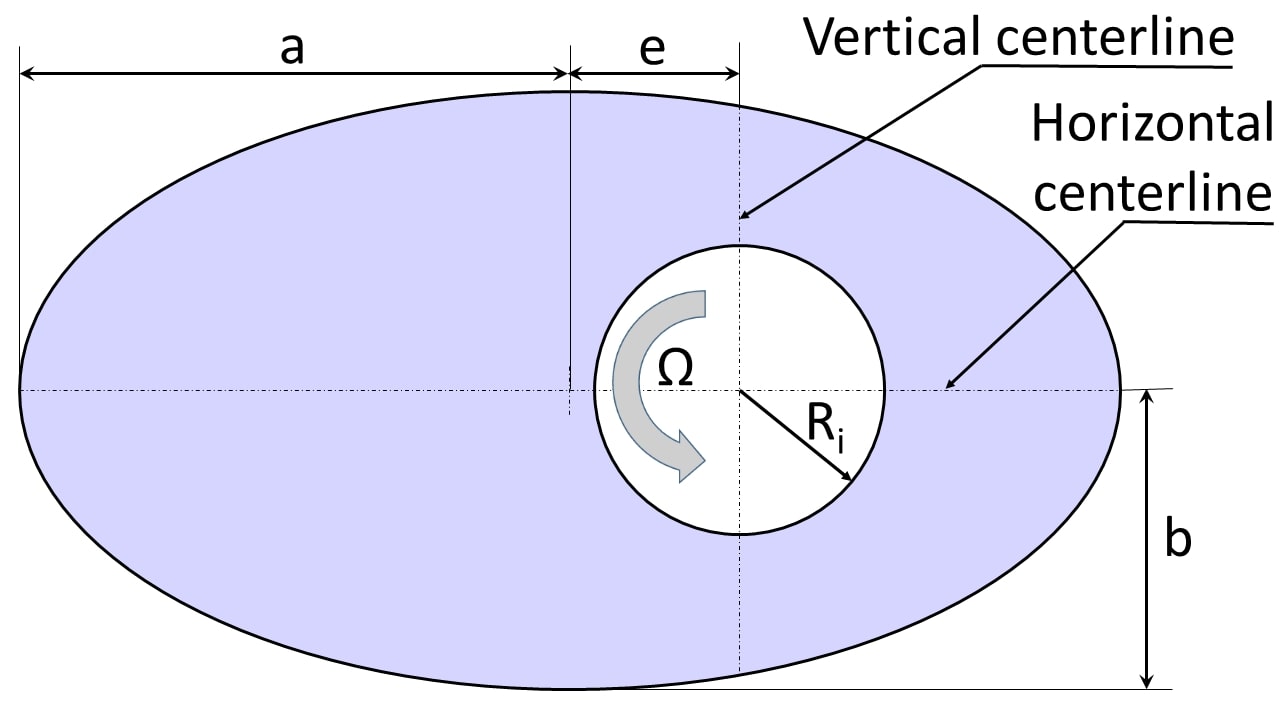}
    \caption{The geometry of the cylinder enclosed within an ellipse}
    \label{fig:elliptical_bearing}
\end{figure}

The intent of this study is to characterize the two-dimensional vortical flow patterns that develop in a stationary elliptical enclosure with a rotating inner cylinder placed concentrically or eccentrically (\cref{fig:elliptical_bearing}). The problem is motivated by the rich fluid physics that is encountered in rotating flows and shear driven flows in a cavity. The formation of multiple vortices in corners and in cavities (known as Moffatt vortices) has been studied previously with one of the bounding walls in motion. The geometry considered here is analogous to the Taylor-Couette flow in circular cylinders but provides richer flow physics by combining the Taylor-Couette flow and Moffatt vortices. We consider different Reynolds numbers $(Re = \omega R_i^2 /\nu)$, eccentricities of the placement of an inner cylinder $(e)$ and different aspect ratios $(l = a/b)$ of the outer elliptic enclosure, where $\omega$ is the angular velocity of the inner cylinder, $R_i$ is the radius of the inner cylinder, $\nu$ is the kinematic viscosity, $e$ is the eccentricity of the placement of the inner cylinder, $a$ is the semi-major axis, and $b$ is the semi-minor axis. For our simulations we have chosen the non-dimensional value of the inner cylinder radius ($R_i$) as 0.5, and the semi minor axis (b) as 1.0. Eccentricity (e) and semi major axis (a) are chosen from the specific eccentricity ratio ($e/a$) and aspect ratio ($a/b$) corresponding to the cases presented in \Cref{sec:results_of_base_case,sec:effect_of_eccentricity,sec:effect_of_aspect_ratio,sec:effect_of_e_and_ar}.

The two-dimensional Navier-Stokes equations in the complex enclosure are solved by a novel high-accuracy meshless method \cite{SHAHANE2021110623} utilizing radial basis function (RBF) interpolations \cite{Bayona2017} of scattered data. We recently developed a semi-implicit time marching algorithm \cite{Shantanu2016} to integrate the time-dependent incompressible fluid flow equations. All the computations in this paper are performed using the open-source Meshless Multi-Physics Software (MeMPhyS) \cite{memphys}.
We use Polyharmonic Splines (PHS) with appended polynomials \cite{SHAHANE2021110623} to achieve high spatial discretization accuracy. Unstructured Finite Volume Methods (FVM) are widely used in commercial fluid flow software but can at best be second-order accurate because of the inherent formulation of the FVM concept unless complex reconstruction schemes are devised. If grid skewness is significant, such methods can even degrade to first-order accuracy. Thus, in problems where highly accurate solutions are required the mesh size has to be really small which adds to the computational cost. The method that is presented here achieves high order convergence of the discretization errors without special reconstruction procedures. The present computations assume the flow to be two-dimensional. However, fully three-dimensional studies will also be conducted in future to determine the onset and characteristics of the three-dimensional flow.

\Cref{sec:numerical_method} describes details of the discretization and the solution algorithm. \Cref{sec:grid_indipendency} presents grid independence calculations to select the appropriate number of scattered points for accurate and efficient computations. In \cref{sec:results_of_base_case}, we first present results of a base case with an aspect ratio (major to minor axis) of the outer enclosure of two and concentric placement of the inner cylinder. The flow is computed for a series of Reynolds numbers and streamlines, vorticity contours, selected velocity profiles, and the torque on the inner cylinder are presented. Velocities are interpolated with high accuracy at selected locations and are provided as benchmark values. \Cref{sec:effect_of_eccentricity,sec:effect_of_aspect_ratio} present the individual effects of eccentric placement of the inner rotating cylinder and the aspect ratio of the outer enclosure respectively. \Cref{sec:effect_of_e_and_ar} presents the combined effects of both eccentricity and aspect ratio. \Cref{sec:summary} summarizes the current findings and future plans to expand this research.

\section{NUMERICAL METHOD}
\label{sec:numerical_method}
\subsection{The PHS-RBF Discretization} \label{Sec:The PHS-RBF Method}
A scalar function $s(\textbf{x})$ such as a velocity component or pressure is interpolated over a set of $q$ nearest neighbors, known as cloud points using the polyharmonic spline radial basis functions (PHS-RBF) appended with monomials:
\begin{equation}
	s(\textbf{x}) = \sum_{i=1}^{q} \lambda_i \phi_i (||\bm{x} - \bm{x_i}||_2) + \sum_{i=1}^{m} \gamma_i P_i (\bm{x})
	\label{Eq:RBF_interp}
\end{equation}
where, $\phi(r)=r^{2a+1},\hspace{0.1cm} a \in \mathbb{N}$ is the PHS-RBF, $m$ is the number of monomials ($P_i$) upto a maximum degree of $k$ and $(\lambda_i, \gamma_i)$ are $q+m$ free parameters. $q$ equations are obtained by collocating \Cref{Eq:RBF_interp} over the $q$ cloud points. Further constraints on the polynomials provide m additional equations required to close the linear system \cite{flyer2016onrole_I}:
\begin{equation}
    \sum_{i=1}^{q} \lambda_i P_j(\bm{x_i}) =0 \hspace{0.5cm} \text{for } 1 \leq j \leq m
\label{Eq:RBF_constraint}
\end{equation}
\Cref{Eq:RBF_constraint} together with the collocation conditions can be written in the matrix vector form:
\begin{equation}
\begin{bmatrix}
\bm{\Phi} & \bm{P}  \\
\bm{P}^T & \bm{0} \\
\end{bmatrix}
\begin{bmatrix}
\bm{\lambda}  \\
\bm{\gamma} \\
\end{bmatrix} =
\begin{bmatrix}
\bm{A}
\end{bmatrix}
\begin{bmatrix}
\bm{\lambda}  \\
\bm{\gamma} \\
\end{bmatrix} =
\begin{bmatrix}
\bm{s}  \\
\bm{0} \\
\end{bmatrix}
\label{Eq:RBF_interp_mat_vec}
\end{equation}
where, transpose is denoted by the superscript $T$, $\bm{\lambda} = [\lambda_1,...,\lambda_q]^T$, $\bm{\gamma} = [\gamma_1,...,\gamma_m]^T$, $\bm{s} = [s(\bm{x_1}),...,s(\bm{x_q})]^T$ and $\bm{0}$ is the matrix of zeros. Dimensions of the submatrices $\bm{P}$ and $\bm{\Phi}$ are $q\times m$ and $q\times q$ respectively.
For a degree of the appended polynomial, say 2 ($k=2$), and for two dimensional problem ($d=2$), there can be $m=\binom{k+d}{k}=6$ polynomial terms, given as $[1, x, y, x^2, xy, y^2]$. The submatrix $\bm{P}$ is evaluated on the $q$ cloud points as given as
\begin{equation}
\bm{P} =
\begin{bmatrix}
1 & x_1  & y_1 & x_1^2 & x_1 y_1 & y_1^2 \\
\vdots & \vdots & \vdots & \vdots & \vdots & \vdots \\
1 & x_q  & y_q & x_q^2 & x_q y_q & y_q^2 \\
\end{bmatrix}
\label{Eq:RBF_interp_poly}
\end{equation}
The submatrix $\bm{\Phi}$ is similarly evaluated at the $q$ cloud points given in \Cref{Eq:RBF_interp_phi}. Here $\phi$ is the PHS-RBF function that we choose.
\begin{equation}
\bm{\Phi} =
\begin{bmatrix}
\phi \left(||\bm{x_1} - \bm{x_1}||_2\right) & \dots  & \phi \left(||\bm{x_1} - \bm{x_q}||_2\right) \\
\vdots & \ddots & \vdots \\
\phi \left(||\bm{x_q} - \bm{x_1}||_2\right) & \dots  & \phi \left(||\bm{x_q} - \bm{x_q}||_2\right) \\
\end{bmatrix}
\label{Eq:RBF_interp_phi}
\end{equation}

Interpolation coefficients $\lambda_i$ and $\gamma_i$ are calculated by solving \Cref{Eq:RBF_interp_mat_vec}:
\begin{equation}
\begin{bmatrix}
\bm{\lambda}  \\
\bm{\gamma} \\
\end{bmatrix} =
\begin{bmatrix}
\bm{A}
\end{bmatrix} ^{-1}
\begin{bmatrix}
\bm{s}  \\
\bm{0} \\
\end{bmatrix}
\label{Eq:RBF_interp_mat_vec_solve}
\end{equation}
Although inverse of the matrix $\bm{A}$ is symbolically shown in \Cref{Eq:RBF_interp_mat_vec_solve}, the dense linear \Cref{Eq:RBF_interp_mat_vec} is solved using a direct solver.
\par The first and second order differential operators in the Navier-Stokes equations are then estimated at any point using a stencil over the neighboring cloud points. Any scalar linear operator $\mathcal{L}$ such as $\frac{\partial}{\partial x}$ or the Laplacian $\nabla ^2$ can be applied in \Cref{Eq:RBF_interp}:
\begin{equation}
\mathcal{L} [s(\textbf{x})] = \sum_{i=1}^{q} \lambda_i \mathcal{L} [\phi_i (\bm{||\bm{x} - \bm{x_i}||_2})] + \sum_{i=1}^{m} \gamma_i \mathcal{L}[P_i (\bm{x})]
\label{Eq:RBF_interp_L}
\end{equation}
Collocating \Cref{Eq:RBF_interp_L} over the $q$ cloud points gives a rectangular matrix vector system:
\begin{equation}
\mathcal{L}[\bm{s}] =
\begin{bmatrix}
\mathcal{L}[\bm{\Phi}] & \mathcal{L}[\bm{P}]  \\
\end{bmatrix}
\begin{bmatrix}
\bm{\lambda}  \\
\bm{\gamma} \\
\end{bmatrix}
\label{Eq:RBF_interp_mat_vec_L}
\end{equation}
where, $\mathcal{L}[\bm{P}]$ and $\mathcal{L}[\bm{\Phi}]$ are matrices of sizes $q\times m$ and $q\times q$ respectively. Substituting \Cref{Eq:RBF_interp_mat_vec_solve} in \Cref{Eq:RBF_interp_mat_vec_L} results in:
\begin{equation}
\begin{aligned}
\mathcal{L}[\bm{s}] &=
\left(\begin{bmatrix}
\mathcal{L}[\bm{\Phi}] & \mathcal{L}[\bm{P}]  \\
\end{bmatrix}
\begin{bmatrix}
\bm{A}
\end{bmatrix} ^{-1}\right)
\begin{bmatrix}
\bm{s}  \\
\bm{0} \\
\end{bmatrix}
=
\begin{bmatrix}
\bm{B}
\end{bmatrix}
\begin{bmatrix}
\bm{s}  \\
\bm{0} \\
\end{bmatrix}\\
&=
\begin{bmatrix}
\bm{B_1} & \bm{B_2}
\end{bmatrix}
\begin{bmatrix}
\bm{s}  \\
\bm{0} \\
\end{bmatrix}
= [\bm{B_1}] [\bm{s}] + [\bm{B_2}] [\bm{0}]
= [\bm{B_1}] [\bm{s}]
\end{aligned}
\label{Eq:RBF_interp_mat_vec_L_solve}
\end{equation}
Multiplying the coefficient matrix $[\bm{B_1}]$ with the discrete field vector $\bm{s}$ gives the numerical estimate of $\mathcal{L}[\bm{s}]$. Collocating the governing equation at a given point gives an implicit equation connecting value at the local point with values at the points in the cloud. Further details of the derivation and properties of the derivative and Laplacian coefficients are given in our previous publications \cite{Shantanu2016,SHAHANE2021110623,bartwal2021application,radhakrishnan2021non}.

\subsection{Solution Algorithm for the Navier-Stokes equation} \label{Sec:Semi-Implicit Algorithm}
We solve the two-dimensional time-dependent Navier-Stokes equations of an incompressible fluid:
\begin{align}
    \nabla . \textbf{u} &= 0\\
    \rho (\frac{\partial \textbf{u}}{\partial t} + \textbf{u}. \nabla \textbf{u}) &= -\nabla p + \mu \nabla^2 \textbf{u}
    \label{EQ:governing equations}
\end{align}
Here $\textbf{u}$ is a vector of Cartesian velocities, $p$ is the pressure, $\mu$ is the fluid viscosity and $t$ is time. Note that the velocities are aligned with the directions $(x, y)$.

In this work, we use the semi-implicit algorithm reported by \citet{Shantanu2016} to solve the Navier-Stokes equations. Let $[u^n, v^n, p^n]$ and $[u^{n-1}, v^{n-1}, p^{n-1}]$ denote the velocity components and pressures at the previous timesteps $n$ and $n-1$. These fields at the next time-step $[u^{n+1}, v^{n+1}, p^{n+1}]$ are determined by iterating. First, the momentum equations are discretized in time using the second-order accurate backward differentiation formula (BDF2) given as:
\begin{equation}
\frac{\rho (\alpha_1 \widetilde{u} + \alpha_2 u^n + \alpha_3 u^{n-1})}{\Delta t} = -\rho \bm{u}^r \bullet (\nabla \widetilde{u}) + \mu \nabla^2 \widetilde{u} - \left(\frac{\partial p}{\partial x}\right)^r
\label{Eq:u:tilde}
\end{equation}
\begin{equation}
\frac{\rho (\alpha_1 \widetilde{v} + \alpha_2 v^n + \alpha_3 v^{n-1})}{\Delta t} = -\rho \bm{u}^r \bullet (\nabla \widetilde{v}) + \mu \nabla^2 \widetilde{v} - \left(\frac{\partial p}{\partial y}\right)^r
\label{Eq:v:tilde}
\end{equation}
where $r$ is the previous iteration number and [$\widetilde{u}$, $\widetilde{v}$] denote intermediate velocity fields. [$\alpha_1, \alpha_2, \alpha_3$] are the BDF2 coefficients given by $[1.5,-2,0.5]$ respectively \cite{suli2003introduction}. The first and second order differential operators are estimated using the PHS-RBF expansions with appended polynomials previously described in \cref{Sec:The PHS-RBF Method}. Intermediate velocity fields $\widetilde{u}$ and $\widetilde{v}$ are obtained by solving \Cref{Eq:u:tilde,Eq:v:tilde} using a preconditioned BiCGSTAB algorithm. Subtracting these equations from the discretized momentum equations and truncating the advection and diffusion terms gives the velocity correction equations:
\begin{equation}
\frac{\rho \alpha_1 (u^{r+1} - \widetilde{u})}{\Delta t} \approx - \frac{\partial (p^{r+1} - p^r)}{\partial x} = - \frac{\partial p'}{\partial x}
\label{Eq:u:p'}
\end{equation}
\begin{equation}
\frac{\rho \alpha_1 (v^{r+1} - \widetilde{v})}{\Delta t} \approx - \frac{\partial (p^{r+1} - p^r)}{\partial y} = - \frac{\partial p'}{\partial y}
\label{Eq:v:p'}
\end{equation}
where, $p'$ denotes corrections to the pressure field. A Poisson equation for pressure corrections is then derived by applying the divergence-free condition on the velocity fields $[u^{r+1}, v^{r+1}]$. This is given by
\begin{equation}
\nabla^2 p' = \frac{\rho \alpha_1}{\Delta t}\left(\frac{\partial \widetilde{u}}{\partial x} + \frac{\partial \widetilde{v}}{\partial y}\right)
\label{Eq:pressure poisson}
\end{equation}
The pressure-correction equation is also discretized to high accuracy using the PHS-RBF expansions and is solved with a preconditioned BiCGSTAB algorithm. The velocities at iteration $r$ are then updated using the pressure correction as:
\begin{equation}
u^{r+1} = \widetilde{u} - \frac{\Delta t}{\rho \alpha_1} \left(\frac{\partial p'}{\partial x}\right)
\label{Eq:u:correction}
\end{equation}
\begin{equation}
v^{r+1} = \widetilde{v} - \frac{\Delta t}{\rho \alpha_1} \left(\frac{\partial p'}{\partial y}\right)
\label{Eq:v:correction}
\end{equation}
Similarly, the pressure is corrected as:
\begin{equation}
p^{r+1} = p^r + p'
\label{Eq:p:correction}
\end{equation}
\par Wall boundary conditions are applied by imposing appropriate velocities in the solution of \Cref{Eq:u:tilde,Eq:v:tilde}. The boundary condition for velocities on ellipse is zero while the velocity at the surface of the inner rotating cylinder is the corresponding tangential velocity of the cylinder (the no-slip and the no penetration conditions). In non-dimensionalised values the tangential velocity is taken as 1.0. The pressure correction \Cref{Eq:pressure poisson} is solved using the homogeneous Neumann boundary condition. The boundary pressures are subsequently estimated from the interior pressures using the normal momentum equation given by:
\begin{equation}
\nabla p \bullet \widehat{\bm{N}} = \left[-\rho\left(\frac{\partial \bm{u}}{\partial t}\right) -\rho(\bm{u}\bullet \nabla)\bm{u} + \mu \nabla^2 \bm{u}\right] \bullet \widehat{\bm{N}}
\label{Eq:normal momentum}
\end{equation}
The above solution steps at any time step can be summarized as:
\begin{enumerate}
	\item First, initialize the velocity and pressure fields at the iteration number $r=0$ as $[u^r, v^r, p^r] = [u^n, v^n, p^n]$
	\item Solve \Cref{Eq:u:tilde,Eq:v:tilde} for intermediate velocities ($\widetilde{u}$ and $\widetilde{v}$) with wall boundary conditions using old pressure field and linearized advection terms
	\item Solve \Cref{Eq:pressure poisson} for pressure correction ($p'$) with homogeneous Neumann boundary conditions
	\item Correct pressures at the interior points (\Cref{Eq:p:correction})
	\item Evaluate the boundary pressures using the normal momentum \Cref{Eq:normal momentum}
	\item Correct the velocities to obtain $u^{r+1}$ and $v^{r+1}$ using the \Cref{Eq:u:correction,Eq:v:correction}
	\item Compute the change in the velocity fields during the current iteration: $\Delta = ||u^{r+1}-u^r|| + ||v^{r+1}-v^r||$
	\item If $\Delta$ is less than the prescribed iterative tolerance $\epsilon$, set $[u^{n+1}, v^{n+1}, p^{n+1}] = [u^{r+1}, v^{r+1}, p^{r+1}]$ and proceed to a new timestep. Otherwise, go back to step 2 and continue the iterations
\end{enumerate}
We initialize the iterative tolerance ($\epsilon$) to  1e-5. Subsequently, it is set as $\min[\epsilon, 0.5 \Delta t \epsilon_s]$ where, $\epsilon_s=\frac{||u^{n+1} - u^n|| + ||v^{n+1} - v^n||}{\Delta t}$ is the steady state error for the velocity fields. Hence, when the solution approaches the steady state, the iterative tolerance is reduced to get a higher convergence accuracy. In order to reduce the computational time, the flow field at a lower Reynolds number is used as an initial guess for a higher Reynolds number. Detailed analysis of the current semi-implicit procedure is presented in \cite{Shantanu2016}.

\section{GRID INDEPENDENCY}
\label{sec:grid_indipendency}

\begin{figure}[h]
    \centering
	\begin{subfigure}[t]{0.45\textwidth}
		\includegraphics[width=\textwidth]{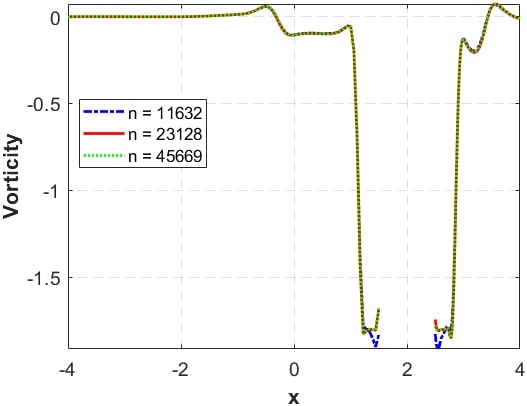}
        \caption{Vorticity vs x}
	\end{subfigure}
	\begin{subfigure}[t]{0.45\textwidth}
		\includegraphics[width=\textwidth]{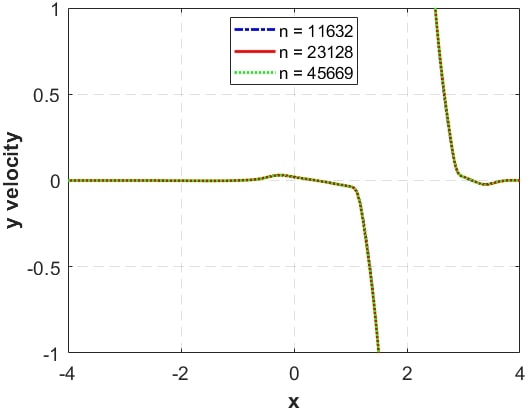}
        \caption{y velocity vs x}
	\end{subfigure}
	\begin{subfigure}[t]{0.45\textwidth}
		\includegraphics[width=\textwidth]{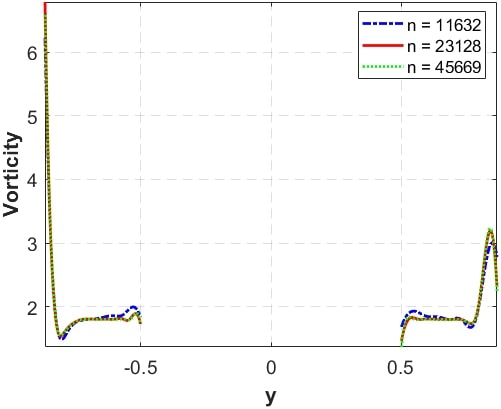}
        \caption{Vorticity vs y}
	\end{subfigure}
	\begin{subfigure}[t]{0.45\textwidth}
		\includegraphics[width=\textwidth]{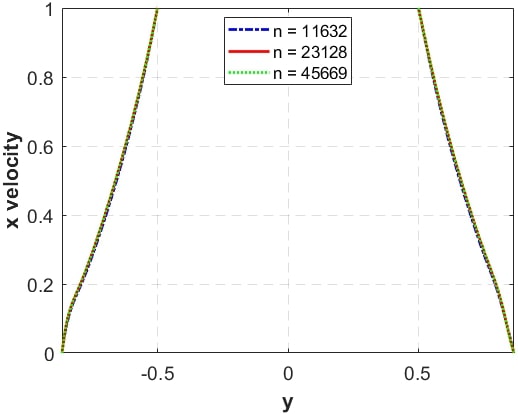}
        \caption{x velocity vs y}
	\end{subfigure}
    \caption{Grid independence study plots : (a) and (b) are plots of vorticity and y velocity with respect to x coordinates along the horizontal centerline respectively; (c) and (d) are plots of vorticity and x velocity with respect to y coordinates along the vertical centerline respectively; In the legends 'n' denotes the number of nodes in each set }
    \label{fig:grid independence}
\end{figure}

For grid independence study, velocity and vorticity plots of three different grids are presented here. We consider the extreme case given by an ellipse with an aspect ratio $(l)$ of 4 and an eccentricity ratio $(e/a)$ of 0.5 for grid independence study. Simulation is run at a Reynolds number of 2000, which is the highest value used in this work. Three grids are selected for grid independence study, with 11632, 23128 and 45669 nodes each. The vertices of the unstructured triangular grid generated with GMSH \cite{Geuzaine2009Gmsh} are used as nodes. In \Cref{fig:grid independence} we present plots of the x-velocity along the vertical line that coincides with the diameter of the inner cylinder (from here on called as the vertical centerline) and the y-velocity along the horizontal line that coincides with the diameter of the inner cylinder (from here on called as the horizontal centerline). The centerlines are shown in \Cref{fig:elliptical_bearing}. It can be observed that their values on the two finer point placements coincide and the coarsest point distribution differs slightly. Vorticity is calculated with derivatives of the velocity components and plotted in \Cref{fig:grid independence} for the same lines. In this case as well, results of the finer point sets coincide with only small differences seen in the coarsest case in regions of steep variation. For best accuracy, we have selected the finest point distribution with an average $\Delta x$ of 1.5E-3 for all calculations presented in this paper. This $\Delta x$ is defined as the average of the distances between the nearest points in all the clouds. Other geometries with different eccentricities and aspect ratios are also discretized with point distributions such that the average $\Delta x$ is around 1.5E-3.

Some of the data that are plotted in \Cref{fig:grid independence} have been tabulated  \Cref{tab:grid_independence_table_1,tab:grid_independence_table_2,tab:grid_independence_table_3,tab:grid_independence_table_4}. The data given in the tables are accurate to the third decimal, and the finest set values and one coarser differ only in the fourth decimal in all locations. For vorticity (which is derived from the computed velocities), the values are accurate to three significant fidures. The points are selected to highlight local maximum and minima. Vorticity given in \Cref{tab:grid_independence_table_1,tab:grid_independence_table_3} ($\eta$) is calculated as
\begin{equation}
    \eta = \frac{\partial v}{\partial x} - \frac{\partial u}{\partial y}
\end{equation}

\begin{table}[H]
\centering
{\begin{tabular}{|c|ccc|}
\hline

Cordinates & \multicolumn{3}{c|}{Vorticity ($\eta$)}                                        \\ \hline
x          & \multicolumn{1}{c|}{n1}      & \multicolumn{1}{c|}{n2}      & n3      \\ \hline
-2.2400    & \multicolumn{1}{c|}{-0.0006} & \multicolumn{1}{c|}{-0.0006} & -0.0005 \\ \hline
-0.4800    & \multicolumn{1}{c|}{0.0575}  & \multicolumn{1}{c|}{0.0599}  & 0.0593  \\ \hline
-0.0400    & \multicolumn{1}{c|}{-0.1056} & \multicolumn{1}{c|}{-0.1079} & -0.1074 \\ \hline
0.2900     & \multicolumn{1}{c|}{-0.0950} & \multicolumn{1}{c|}{-0.0971} & -0.0969 \\ \hline
0.4000     & \multicolumn{1}{c|}{-0.0956} & \multicolumn{1}{c|}{-0.0976} & -0.0974 \\ \hline
0.5650     & \multicolumn{1}{c|}{-0.0968} & \multicolumn{1}{c|}{-0.0989} & -0.0987 \\ \hline
0.9500     & \multicolumn{1}{c|}{-0.0517} & \multicolumn{1}{c|}{-0.0529} & -0.0527 \\ \hline
1.2250     & \multicolumn{1}{c|}{-1.7860} & \multicolumn{1}{c|}{-1.8299} & -1.8285 \\ \hline
1.3350     & \multicolumn{1}{c|}{-1.8138} & \multicolumn{1}{c|}{-1.8014} & -1.7999 \\ \hline
1.4450     & \multicolumn{1}{c|}{-1.9022} & \multicolumn{1}{c|}{-1.8065} & -1.8059 \\ \hline
1.5000     & \multicolumn{1}{c|}{-1.8328} & \multicolumn{1}{c|}{-1.6858} & -1.6807 \\ \hline
2.5000     & \multicolumn{1}{c|}{-1.8280} & \multicolumn{1}{c|}{-1.7448} & -1.7793 \\ \hline
2.5450     & \multicolumn{1}{c|}{-1.9147} & \multicolumn{1}{c|}{-1.8089} & -1.8075 \\ \hline
2.6200     & \multicolumn{1}{c|}{-1.8432} & \multicolumn{1}{c|}{-1.8026} & -1.8027 \\ \hline
2.7100     & \multicolumn{1}{c|}{-1.7826} & \multicolumn{1}{c|}{-1.7932} & -1.7923 \\ \hline
2.7700     & \multicolumn{1}{c|}{-1.8159} & \multicolumn{1}{c|}{-1.8510} & -1.8495 \\ \hline
2.9950     & \multicolumn{1}{c|}{-0.1278} & \multicolumn{1}{c|}{-0.1315} & -0.1313 \\ \hline
3.0700     & \multicolumn{1}{c|}{-0.1632} & \multicolumn{1}{c|}{-0.1692} & -0.1682 \\ \hline
3.5200     & \multicolumn{1}{c|}{0.0641}  & \multicolumn{1}{c|}{0.0662}  & 0.0670  \\ \hline
\end{tabular}%
}
\caption{Grid independence vorticity data at selected locations on the three sets of point placements where, n1 = 11632, n2 = 23128 and n3 = 45669, along the horizontal centerline (y =0.0). Eccentricity ratio for the simulation data presented here is 0.5, the aspect ratio is 4 and the Reynolds number chosen is 2000}
\label{tab:grid_independence_table_1}
\end{table}

\begin{table}[H]
\centering
{\begin{tabular}{|c|ccc|}
\hline
Cordinates & \multicolumn{3}{c|}{y velocity}                                       \\ \hline
x          & \multicolumn{1}{c|}{n1}      & \multicolumn{1}{c|}{n2}      & n3      \\ \hline
-1.8000    & \multicolumn{1}{c|}{-0.0014} & \multicolumn{1}{c|}{-0.0014} & -0.0013 \\ \hline
-1.4700    & \multicolumn{1}{c|}{-0.0018} & \multicolumn{1}{c|}{-0.0018} & -0.0018 \\ \hline
-1.2500    & \multicolumn{1}{c|}{-0.0014} & \multicolumn{1}{c|}{-0.0014} & -0.0014 \\ \hline
-0.7000    & \multicolumn{1}{c|}{0.0057}  & \multicolumn{1}{c|}{0.0057}  & 0.0053  \\ \hline
-0.2600    & \multicolumn{1}{c|}{0.0304}  & \multicolumn{1}{c|}{0.0312}  & 0.0308  \\ \hline
-0.1500    & \multicolumn{1}{c|}{0.0281}  & \multicolumn{1}{c|}{0.0288}  & 0.0289  \\ \hline
0.9500     & \multicolumn{1}{c|}{-0.0317} & \multicolumn{1}{c|}{-0.0325} & -0.0322 \\ \hline
2.6200     & \multicolumn{1}{c|}{0.6026}  & \multicolumn{1}{c|}{0.6115}  & 0.6112  \\ \hline
2.7700     & \multicolumn{1}{c|}{0.2410}  & \multicolumn{1}{c|}{0.2480}  & 0.2480  \\ \hline
2.9200     & \multicolumn{1}{c|}{0.0364}  & \multicolumn{1}{c|}{0.0376}  & 0.0373  \\ \hline
3.0700     & \multicolumn{1}{c|}{0.0116}  & \multicolumn{1}{c|}{0.0124}  & 0.0119  \\ \hline
3.2200     & \multicolumn{1}{c|}{-0.0093} & \multicolumn{1}{c|}{-0.0090} & -0.0094 \\ \hline
3.3700     & \multicolumn{1}{c|}{-0.0224} & \multicolumn{1}{c|}{-0.0233} & -0.0231 \\ \hline
3.3850     & \multicolumn{1}{c|}{-0.0226} & \multicolumn{1}{c|}{-0.0236} & -0.0233 \\ \hline
3.5200     & \multicolumn{1}{c|}{-0.0149} & \multicolumn{1}{c|}{-0.0160} & -0.0153 \\ \hline
3.6700     & \multicolumn{1}{c|}{-0.0034} & \multicolumn{1}{c|}{-0.0037} & -0.0035 \\ \hline
3.8650     & \multicolumn{1}{c|}{0.0009}  & \multicolumn{1}{c|}{0.0009}  & 0.0009  \\ \hline
\end{tabular}%
}
\caption{Grid independence y-velocity data at selected locations on the three sets of point placements where, n1 = 11632, n2 = 23128 and n3 = 45669, along the horizontal centerline (y = 0.0). Eccentricity ratio for the simulation data presented here is 0.5, the aspect ratio is 4 and the Reynolds number chosen is 2000}
\label{tab:grid_independence_table_2}
\end{table}

\begin{table}[H]
\centering
{\begin{tabular}{|c|ccc|}
\hline
Cordinates & \multicolumn{3}{c|}{Vorticity ($\eta$)}                                        \\ \hline
y          & \multicolumn{1}{c|}{n1}      & \multicolumn{1}{c|}{n2}      & n3      \\ \hline
-0.8624    & \multicolumn{1}{c|}{-5.5522} & \multicolumn{1}{c|}{-5.9607} & -5.8612 \\ \hline
-0.8294    & \multicolumn{1}{c|}{-2.0284} & \multicolumn{1}{c|}{-1.9826} & -1.9791 \\ \hline
-0.8075    & \multicolumn{1}{c|}{-1.4987} & \multicolumn{1}{c|}{-1.5393} & -1.5447 \\ \hline
-0.7562    & \multicolumn{1}{c|}{-1.7306} & \multicolumn{1}{c|}{-1.7459} & -1.7443 \\ \hline
-0.6977    & \multicolumn{1}{c|}{-1.8036} & \multicolumn{1}{c|}{-1.8064} & -1.8038 \\ \hline
-0.6537    & \multicolumn{1}{c|}{-1.8240} & \multicolumn{1}{c|}{-1.8023} & -1.8015 \\ \hline
-0.6098    & \multicolumn{1}{c|}{-1.8580} & \multicolumn{1}{c|}{-1.8004} & -1.8031 \\ \hline
-0.5988    & \multicolumn{1}{c|}{-1.8568} & \multicolumn{1}{c|}{-1.8044} & -1.8030 \\ \hline
-0.5695    & \multicolumn{1}{c|}{-1.8666} & \multicolumn{1}{c|}{-1.8107} & -1.8069 \\ \hline
-0.5476    & \multicolumn{1}{c|}{-1.9463} & \multicolumn{1}{c|}{-1.7861} & -1.7924 \\ \hline
-0.5146    & \multicolumn{1}{c|}{-1.9611} & \multicolumn{1}{c|}{-1.8864} & -1.9024 \\ \hline
-0.5000    & \multicolumn{1}{c|}{-1.8333} & \multicolumn{1}{c|}{-1.7274} & -1.7289 \\ \hline
0.5293     & \multicolumn{1}{c|}{-1.9022} & \multicolumn{1}{c|}{-1.8141} & -1.8405 \\ \hline
0.5586     & \multicolumn{1}{c|}{-1.9226} & \multicolumn{1}{c|}{-1.8098} & -1.7996 \\ \hline
0.5805     & \multicolumn{1}{c|}{-1.8765} & \multicolumn{1}{c|}{-1.8009} & -1.8047 \\ \hline
0.6025     & \multicolumn{1}{c|}{-1.8446} & \multicolumn{1}{c|}{-1.8072} & -1.8033 \\ \hline
0.6720     & \multicolumn{1}{c|}{-1.8053} & \multicolumn{1}{c|}{-1.8005} & -1.7997 \\ \hline
0.6903     & \multicolumn{1}{c|}{-1.7997} & \multicolumn{1}{c|}{-1.8008} & -1.7999 \\ \hline
0.7672     & \multicolumn{1}{c|}{-1.6756} & \multicolumn{1}{c|}{-1.7323} & -1.7362 \\ \hline
0.8404     & \multicolumn{1}{c|}{-2.9738} & \multicolumn{1}{c|}{-3.1964} & -3.2321 \\ \hline
\end{tabular}%
}
\caption{Grid independence vorticity data at selected locations on the three sets of point placements where, n1 = 11632, n2 = 23128 and n3 = 45669, along the vertical centerline (x = 2.0). Eccentricity ratio for the simulation data presented here is 0.5, the aspect ratio is 4 and the Reynolds number chosen is 2000.}
\label{tab:grid_independence_table_3}
\end{table}

\begin{table}[H]
\centering
{\begin{tabular}{|c|ccc|}
\hline
Cordinates & \multicolumn{3}{c|}{x velocity}                                       \\ \hline
y          & \multicolumn{1}{c|}{n1}      & \multicolumn{1}{c|}{n2}      & n3      \\ \hline
-0.8624    & \multicolumn{1}{c|}{0.0206}  & \multicolumn{1}{c|}{0.0225}  & 0.0222  \\ \hline
-0.8294    & \multicolumn{1}{c|}{0.1329}  & \multicolumn{1}{c|}{0.1384}  & 0.1386  \\ \hline
-0.7928    & \multicolumn{1}{c|}{0.1986}  & \multicolumn{1}{c|}{0.2050}  & 0.2053  \\ \hline
-0.7562    & \multicolumn{1}{c|}{0.2695}  & \multicolumn{1}{c|}{0.2776}  & 0.2777  \\ \hline
-0.7196    & \multicolumn{1}{c|}{0.3489}  & \multicolumn{1}{c|}{0.3583}  & 0.3584  \\ \hline
-0.6830    & \multicolumn{1}{c|}{0.4353}  & \multicolumn{1}{c|}{0.4452}  & 0.4452  \\ \hline
-0.6464    & \multicolumn{1}{c|}{0.5280}  & \multicolumn{1}{c|}{0.5382}  & 0.5381  \\ \hline
-0.6098    & \multicolumn{1}{c|}{0.6293}  & \multicolumn{1}{c|}{0.6384}  & 0.6382  \\ \hline
-0.5732    & \multicolumn{1}{c|}{0.7397}  & \multicolumn{1}{c|}{0.7472}  & 0.7471  \\ \hline
-0.5366    & \multicolumn{1}{c|}{0.8624}  & \multicolumn{1}{c|}{0.8662}  & 0.8661  \\ \hline
0.5293     & \multicolumn{1}{c|}{-0.8929} & \multicolumn{1}{c|}{-0.8965} & -0.8961 \\ \hline
0.5659     & \multicolumn{1}{c|}{-0.7673} & \multicolumn{1}{c|}{-0.7744} & -0.7742 \\ \hline
0.6025     & \multicolumn{1}{c|}{-0.6547} & \multicolumn{1}{c|}{-0.6638} & -0.6636 \\ \hline
0.6391     & \multicolumn{1}{c|}{-0.5522} & \multicolumn{1}{c|}{-0.5623} & -0.5621 \\ \hline
0.6757     & \multicolumn{1}{c|}{-0.4583} & \multicolumn{1}{c|}{-0.4684} & -0.4683 \\ \hline
0.7123     & \multicolumn{1}{c|}{-0.3715} & \multicolumn{1}{c|}{-0.3811} & -0.3810 \\ \hline
0.7489     & \multicolumn{1}{c|}{-0.2904} & \multicolumn{1}{c|}{-0.2995} & -0.2996 \\ \hline
0.7855     & \multicolumn{1}{c|}{-0.2171} & \multicolumn{1}{c|}{-0.2242} & -0.2242 \\ \hline
0.8221     & \multicolumn{1}{c|}{-0.1299} & \multicolumn{1}{c|}{-0.1339} & -0.1337 \\ \hline
0.8587     & \multicolumn{1}{c|}{-0.0212} & \multicolumn{1}{c|}{-0.0189} & -0.0182 \\ \hline
\end{tabular}%
}
\caption{Grid independence x-velocity data at selected locations on the three sets of point placements where, n1 = 11632, n2 = 23128 and n3 = 45669, along the vertical centerline (x = 2.0). Eccentricity ratio for the simulation data presented here is 0.5, the aspect ratio is 4 and the Reynolds number chosen is 2000}
\label{tab:grid_independence_table_4}
\end{table}

\section{RESULTS OF BASE CASE}
\label{sec:results_of_base_case}
\begin{figure}[h]
    \centering
	\begin{subfigure}[t]{0.45\textwidth}
		\includegraphics[width=\textwidth]{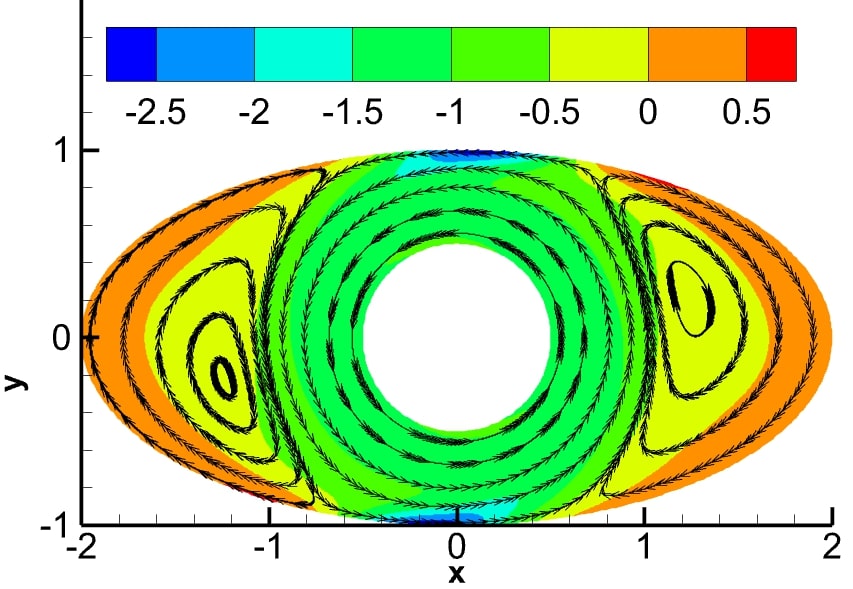}
        \caption{}
        \label{fig:case1a-a}
	\end{subfigure}
	\begin{subfigure}[t]{0.45\textwidth}
		\includegraphics[width=\textwidth]{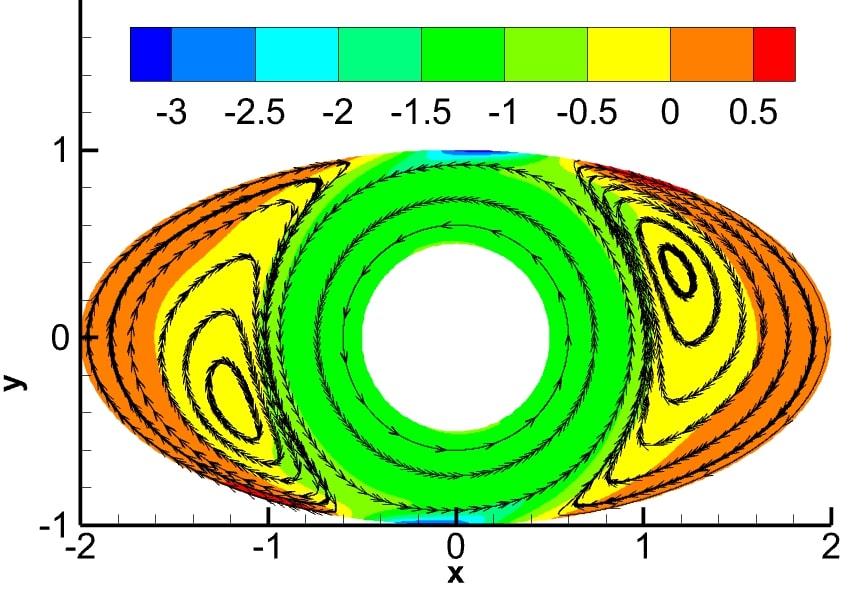}
        \caption{}
        \label{fig:case1a-b}
	\end{subfigure}
	\begin{subfigure}[t]{0.45\textwidth}
		\includegraphics[width=\textwidth]{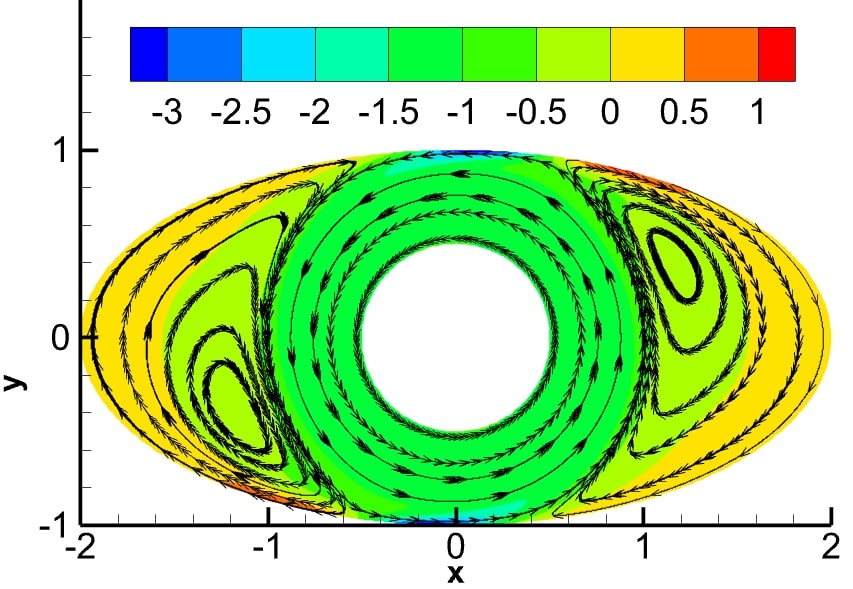}
        \caption{}
        \label{fig:case1a-c}
	\end{subfigure}
	\begin{subfigure}[t]{0.45\textwidth}
		\includegraphics[width=\textwidth]{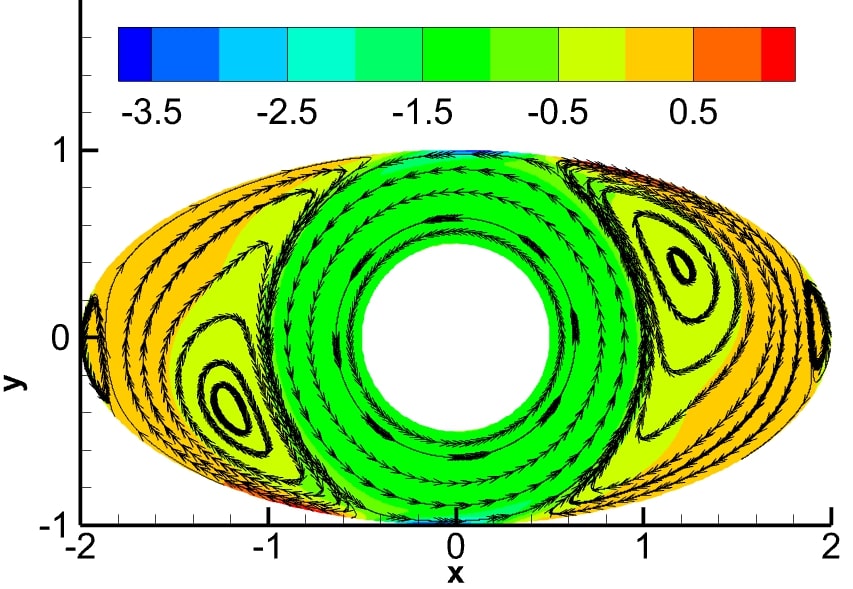}
        \caption{}
        \label{fig:case1a-d}
	\end{subfigure}
	\caption{Streamlines and vorticity contours for the case with aspect ratio is 2 and no eccentricity at four Reynolds numbers a) $Re=200$, b)$Re = 600$, c) $Re = 1000$ and d) $Re = 2000$}
    \label{fig:case1a}
\end{figure}

In this section, we analyze the flow between the ellipse with an aspect ratio of 2 and the circular cylinder placed at the center symmetrically (zero eccentricity). This is defined as the base case and the effects of the aspect ratio and eccentricity over the base case are presented systematically in \cref{sec:effect_of_eccentricity,sec:effect_of_aspect_ratio,sec:effect_of_e_and_ar}. For this base case, the Reynolds number is varied from 200 to 2000 with an increase of 200. To simulate higher Reynolds numbers, we
decreased the viscosity, keeping the tangential velocity of the inner cylinder constant. \Cref{fig:case1a} shows vorticity contours with superimposed streamlines for four Reynolds numbers. Two distinct features can be noted in these figures.

\begin{figure}[H]
    \centering
	\includegraphics[width=0.5\textwidth]{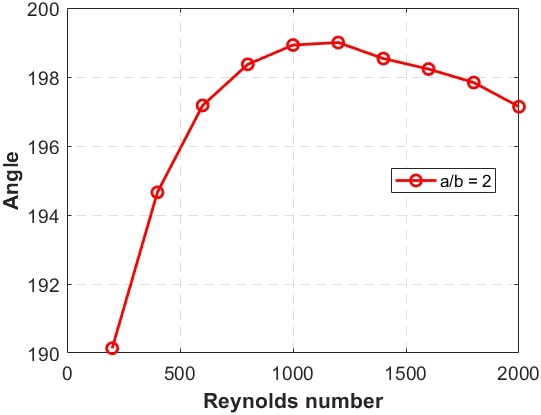}
    \caption{Plot of angle of the primary vortex vs Reynolds number for the case with an aspect ratio of 2 and no eccentricity}
    \vspace{0.25cm}
    \label{fig:angle_case1a}
\end{figure}

First, at lower Reynolds numbers, the center of vortices shifts in the counterclockwise direction of the rotating cylinder. The angle and distance of eye of primary vortices with respect to the horizontal centerline increase with an increase in the Reynolds number. This angle increases and reaches a maximum. When the Reynolds number is around 1000, the angle made by the vortex center decreases as the Reynolds number is increased. This behavior is shown in \Cref{fig:angle_case1a}

As the Reynolds number is increased from 200 to 2000, we also see the formation of secondary vortices. These vortices are similar to the Moffatt vortices that have been observed in sharp corners such as in triangular cavities. However, these are smaller in size and are formed at the most slender portion, that is, along the major axis of the ellipse. The secondary vortices that are formed at higher Reynolds numbers are also tilted in the direction of rotation of the inner cylinder with respect to the horizontal centerline. Benchmark data in \Cref{tab:case 1a x} provides the vorticity ($\eta$) and y - velocity data along the horizontal centerline and \Cref{tab:case 1a y} contains the vorticity and x - velocity data along the vertical centerline for three different Reynolds numbers of 400, 1000 and 2000.

\begin{table}[H]
\centering
\begin{tabular}{|c|cc|cc|cc|}
\hline
Cordinates & \multicolumn{2}{c|}{Re = 400}                      & \multicolumn{2}{c|}{Re = 1000}                     & \multicolumn{2}{c|}{Re = 2000}                     \\ \hline
x          & \multicolumn{1}{c|}{v}       & $\eta$ & \multicolumn{1}{c|}{v}       & $\eta$ & \multicolumn{1}{c|}{v}       & $\eta$ \\ \hline
-1.4300    & \multicolumn{1}{c|}{0.0163}  & -0.0769             & \multicolumn{1}{c|}{0.0129}  & -0.0476             & \multicolumn{1}{c|}{0.0124}  & -0.0323             \\ \hline
-1.1450    & \multicolumn{1}{c|}{-0.0240} & -0.2899             & \multicolumn{1}{c|}{-0.0137} & -0.1888             & \multicolumn{1}{c|}{-0.0100} & -0.1093             \\ \hline
-0.8600    & \multicolumn{1}{c|}{-0.2583} & -1.1279             & \multicolumn{1}{c|}{-0.2505} & -1.1884             & \multicolumn{1}{c|}{-0.2450} & -1.1981             \\ \hline
-0.5750    & \multicolumn{1}{c|}{-0.7895} & -1.1134             & \multicolumn{1}{c|}{-0.7884} & -1.1604             & \multicolumn{1}{c|}{-0.7871} & -1.1823             \\ \hline
0.5000     & \multicolumn{1}{c|}{1.0000}  & -1.2376             & \multicolumn{1}{c|}{1.0000}  & -1.1965             & \multicolumn{1}{c|}{1.0000}  & -1.1734             \\ \hline
0.7850     & \multicolumn{1}{c|}{0.3730}  & -1.1518             & \multicolumn{1}{c|}{0.3667}  & -1.1613             & \multicolumn{1}{c|}{0.3615}  & -1.1823             \\ \hline
1.0700     & \multicolumn{1}{c|}{0.0539}  & -0.4667             & \multicolumn{1}{c|}{0.0367}  & -0.4045             & \multicolumn{1}{c|}{0.0264}  & -0.3333             \\ \hline
1.3550     & \multicolumn{1}{c|}{-0.0115} & -0.1103             & \multicolumn{1}{c|}{-0.0093} & -0.0719             & \multicolumn{1}{c|}{-0.0087} & -0.0649             \\ \hline
1.6400     & \multicolumn{1}{c|}{-0.0162} & 0.0024              & \multicolumn{1}{c|}{-0.0112} & 0.0184              & \multicolumn{1}{c|}{-0.0083} & 0.0394              \\ \hline
1.9250     & \multicolumn{1}{c|}{-0.0032} & 0.0384              & \multicolumn{1}{c|}{-0.0014} & 0.0198              & \multicolumn{1}{c|}{0.0000}  & 0.0055              \\ \hline
\end{tabular}%
\caption{Benchmark data for a/b = 2, y = 0.0 and e/a = 0.0, along the horizontal centerline}
\label{tab:case 1a x}
\end{table}

\begin{table}[H]
\centering
\begin{tabular}{|c|cc|cc|cc|}
\hline
Cordinates & \multicolumn{2}{c|}{Re = 400}                      & \multicolumn{2}{c|}{Re = 1000}                     & \multicolumn{2}{c|}{Re = 2000}                     \\ \hline
y          & \multicolumn{1}{c|}{u}       & $\eta$ & \multicolumn{1}{c|}{u}       & $\eta$ & \multicolumn{1}{c|}{u}       & $\eta$ \\ \hline
-0.9750    & \multicolumn{1}{c|}{0.0685}  & -2.5331             & \multicolumn{1}{c|}{0.0723}  & -2.5682             & \multicolumn{1}{c|}{0.0746}  & -2.5119             \\ \hline
-0.8800    & \multicolumn{1}{c|}{0.2480}  & -1.1955             & \multicolumn{1}{c|}{0.2323}  & -1.0927             & \multicolumn{1}{c|}{0.2213}  & -1.1433             \\ \hline
-0.7850    & \multicolumn{1}{c|}{0.3829}  & -1.0965             & \multicolumn{1}{c|}{0.3708}  & -1.1603             & \multicolumn{1}{c|}{0.3640}  & -1.1839             \\ \hline
-0.6900    & \multicolumn{1}{c|}{0.5457}  & -1.1297             & \multicolumn{1}{c|}{0.5375}  & -1.1604             & \multicolumn{1}{c|}{0.5327}  & -1.1826             \\ \hline
-0.5950    & \multicolumn{1}{c|}{0.7465}  & -1.1373             & \multicolumn{1}{c|}{0.7410}  & -1.1600             & \multicolumn{1}{c|}{0.7381}  & -1.1819             \\ \hline
-0.5000    & \multicolumn{1}{c|}{1.0000}  & -0.5938             & \multicolumn{1}{c|}{1.0000}  & -0.7917             & \multicolumn{1}{c|}{1.0000}  & -0.9697             \\ \hline
0.5000     & \multicolumn{1}{c|}{-1.0000} & -0.5990             & \multicolumn{1}{c|}{-1.0000} & -0.7886             & \multicolumn{1}{c|}{-1.0000} & -0.9533             \\ \hline
0.5950     & \multicolumn{1}{c|}{-0.7465} & -1.1370             & \multicolumn{1}{c|}{-0.7410} & -1.1597             & \multicolumn{1}{c|}{-0.7381} & -1.1824             \\ \hline
0.6900     & \multicolumn{1}{c|}{-0.5457} & -1.1298             & \multicolumn{1}{c|}{-0.5375} & -1.1605             & \multicolumn{1}{c|}{-0.5327} & -1.1827             \\ \hline
0.7850     & \multicolumn{1}{c|}{-0.3829} & -1.0965             & \multicolumn{1}{c|}{-0.3708} & -1.1603             & \multicolumn{1}{c|}{-0.3640} & -1.1839             \\ \hline
0.8800     & \multicolumn{1}{c|}{-0.2480} & -1.1954             & \multicolumn{1}{c|}{-0.2323} & -1.0926             & \multicolumn{1}{c|}{-0.2213} & -1.1433             \\ \hline
0.9750     & \multicolumn{1}{c|}{-0.0685} & -2.5331             & \multicolumn{1}{c|}{-0.0723} & -2.5680             & \multicolumn{1}{c|}{-0.0746} & -2.5111             \\ \hline
\end{tabular}%
\caption{Benchmark data for a/b = 2, x = 0.0 and e/a = 0.0, along the vertical centerline}
\label{tab:case 1a y}
\end{table}

\section{EFFECT  OF ECCENTRICITY}
\label{sec:effect_of_eccentricity}
We first consider the effect of placing the inner cylinder eccentrically within the outer ellipse. Two eccentricity ratios of 0.25 and 0.5 of the semi major axis of the ellipse are considered. The eccentricity ratios remain constant for all aspect ratios. In this section, we first present results for the base case with an aspect ratio of 2.

\begin{figure}[t]
	\centering
	\begin{subfigure}[t]{0.45\textwidth}
		\includegraphics[width=\textwidth]{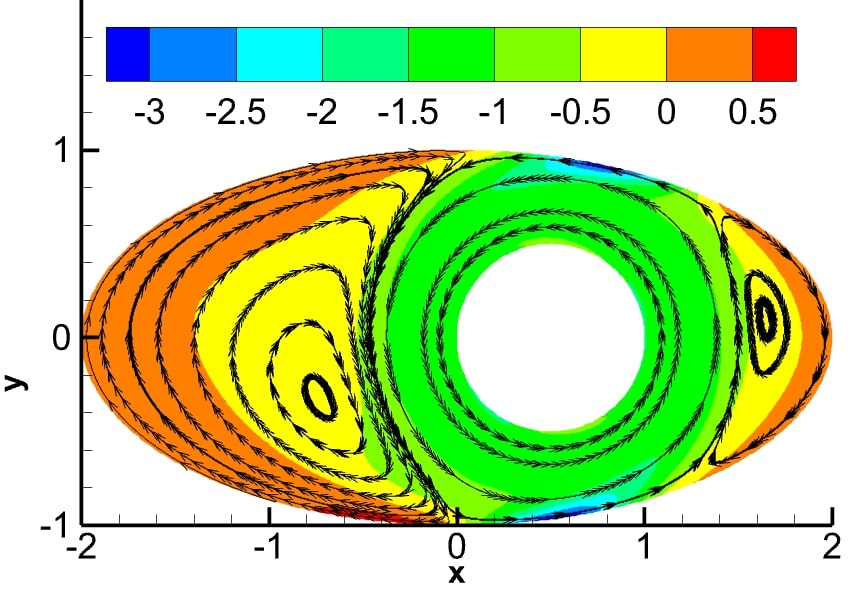}
        \caption{}
        \vspace{0.25cm}
        \label{fig:case1b-a}
	\end{subfigure}
	\begin{subfigure}[t]{0.45\textwidth}
		\includegraphics[width=\textwidth]{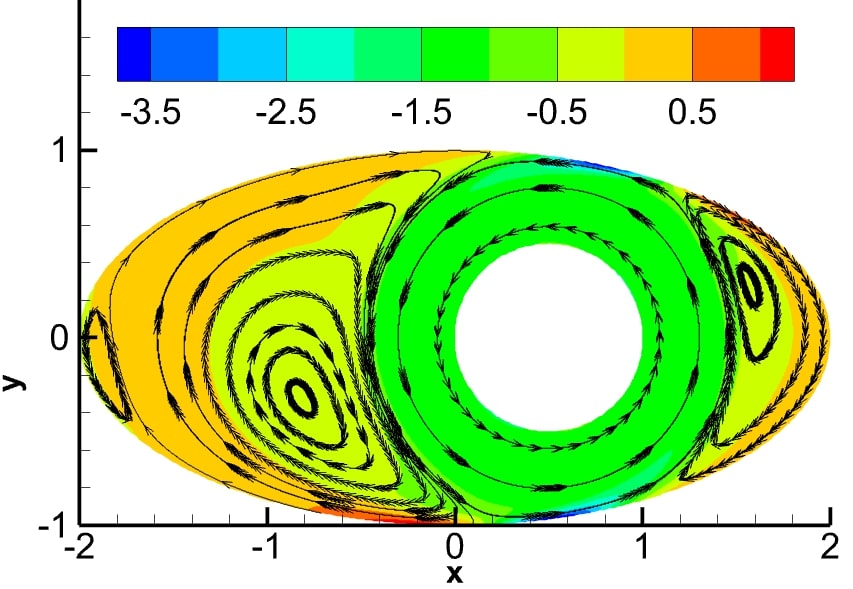}
        \caption{}
        \vspace{0.25cm}
        \label{fig:case1b-b}
	\end{subfigure}
	\begin{subfigure}[t]{0.45\textwidth}
		\includegraphics[width=\textwidth]{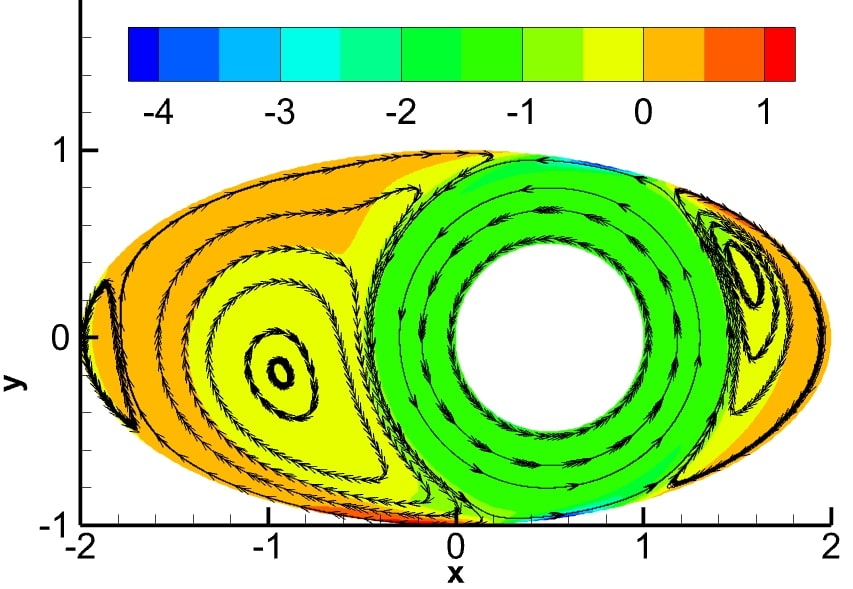}
        \caption{}
        \vspace{0.25cm}
        \label{fig:case1b-c}
	\end{subfigure}
	\caption{Streamlines and vorticity contours for the case where the aspect ratio is 2 and the eccentricity ratio is 0.25 at three Reynolds numbers a) $Re=200$, b) $Re = 1000$ and c) $Re = 2000$}
    \label{fig:case1b}
\end{figure}

\begin{figure}[h]
	\centering
	\begin{subfigure}[t]{0.45\textwidth}
		\includegraphics[width=\textwidth]{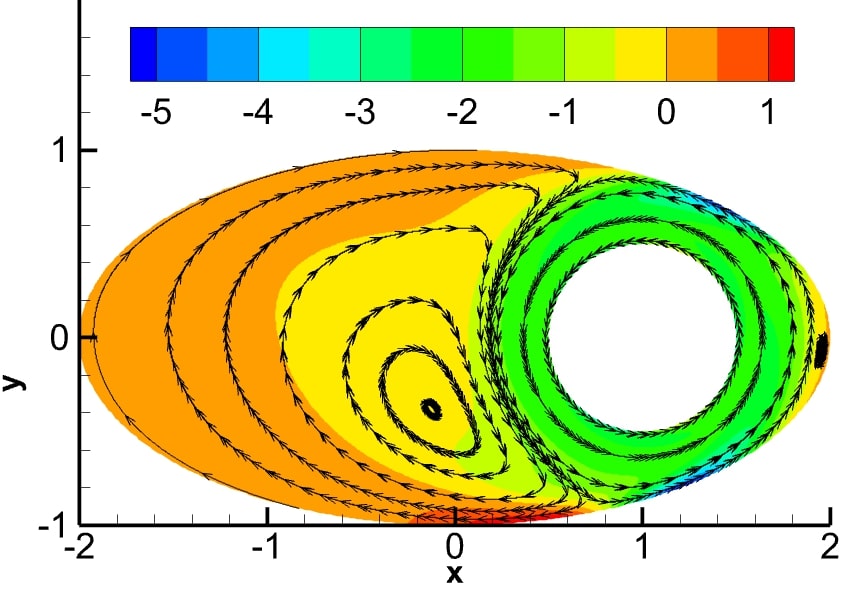}
        \caption{}
        \vspace{0.25cm}
	\end{subfigure}
	\begin{subfigure}[t]{0.45\textwidth}
		\includegraphics[width=\textwidth]{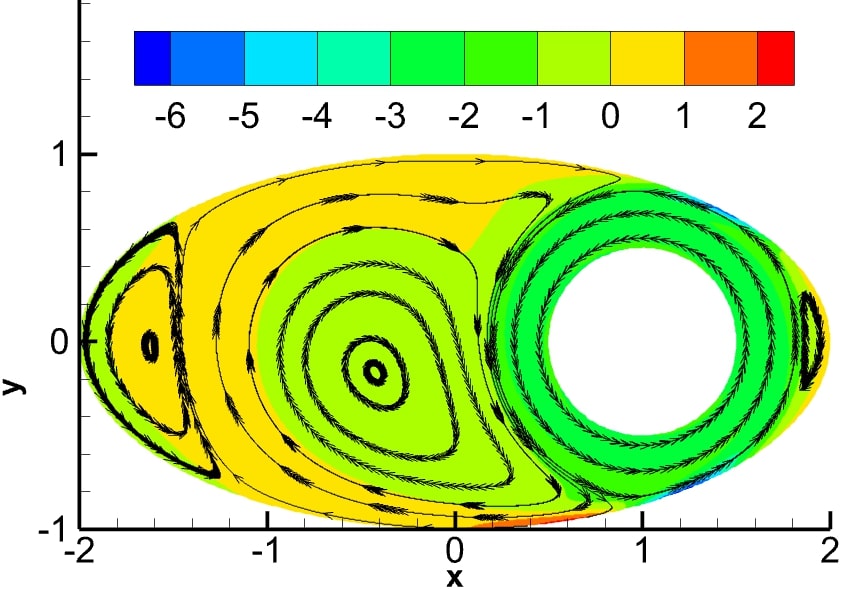}
        \caption{}
        \vspace{0.25cm}
        \end{subfigure}
	\begin{subfigure}[t]{0.45\textwidth}
		\includegraphics[width=\textwidth]{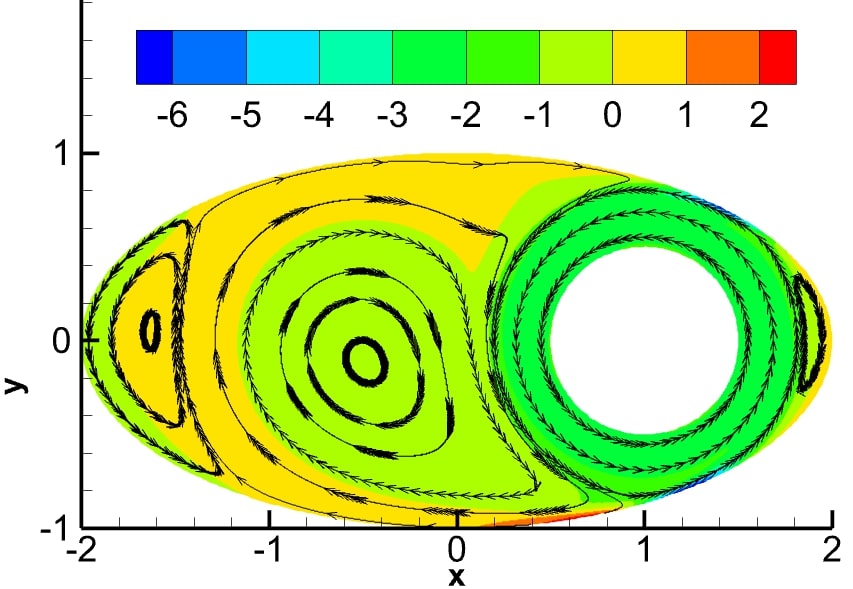}
        \caption{}
        \vspace{0.25cm}
	\end{subfigure}
	\caption{Streamlines and vorticity contour for the case where the aspect ratio is 2 and the eccentricity ratio is 0.5 at three Reynolds numbers a) $Re=200$, b) $Re = 1000$ and c) $Re = 2000$}
    \label{fig:case1c}
\end{figure}

\Cref{fig:case1b,fig:case1c} show the effect of eccentricity on the flow structures. We first observe that the secondary vortex on the smaller gap side disappears for the two eccentricities for all the Reynolds numbers. Moreover the secondary vortex appears at a lower Reynolds number of 1000 on the wide gap side as shown in \Cref{fig:case1b-b} in contrast to the base case of \Cref{fig:case1a-c}. As the Reynolds number is increased, the eye of the primary vortex moves toward the horizontal centerline on the wide gap region while the eye of vortex on the short gap side moves away from the horizontal centerline.

As the eccentricity ratio is increased to 0.5, the primary vortex on the short gap side becomes much smaller, while the one on the wide gap side grows in size as shown in \Cref{fig:case1c}. A bigger secondary vortex is formed on the wide gap side at a lower Reynolds number than the case with lower eccentricity. Secondary vortices grow in size with Reynolds number. It is also notable that the streamlines around the cylinder become more circular as the Reynolds number is increased.

\begin{figure}[H]
    \centering
        \includegraphics[width = 0.45\textwidth]{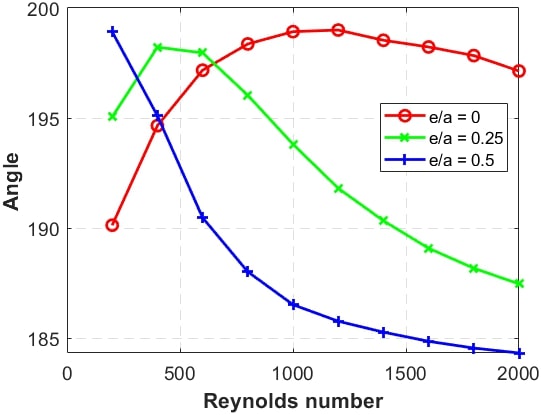}

    \caption{Plot of the primary vortex eye angle with respect to the Reynolds number}
    \label{fig:angle_case1}
\end{figure}

The effect of eccentricity on the angle of the eye of the primary vortex on the major gap side is depicted in \Cref{fig:angle_case1}. It can be observed that for the base case of zero eccentricity the trend is that the magnitude of the angle first increases and then decreases as the Reynolds number is increased. For the smaller eccentricity, the Reynolds number at which the angle peaks is reduced. Finally, for the highest eccentricity that has been simulated, the angle tends to continuously decrease as the Reynolds number is increased.

\section{EFFECT OF ASPECT RATIO}
\label{sec:effect_of_aspect_ratio}
\begin{figure}[H]
	\centering
	\begin{subfigure}[t]{0.45\textwidth}
		\includegraphics[width=\textwidth]{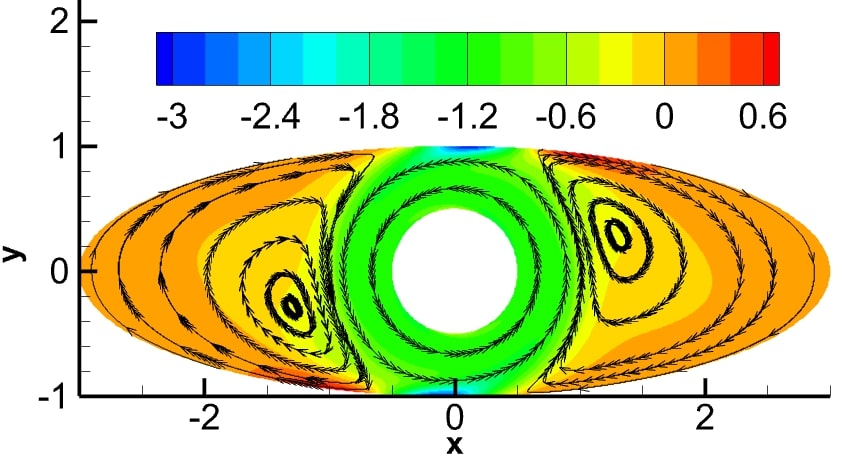}
        \caption{}
        \vspace{0.25cm}
	\end{subfigure}
	\begin{subfigure}[t]{0.45\textwidth}
		\includegraphics[width=\textwidth]{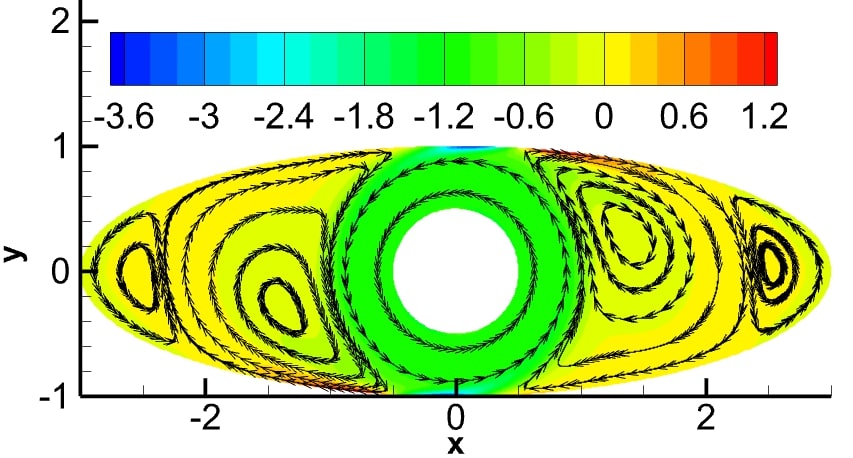}
        \caption{}
        \vspace{0.25cm}
	\end{subfigure}
	\begin{subfigure}[t]{0.45\textwidth}
		\includegraphics[width=\textwidth]{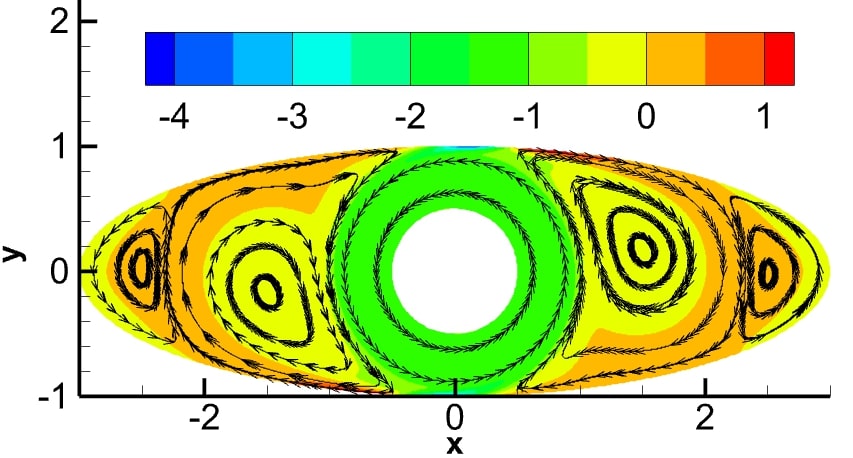}
        \caption{}
        \vspace{0.25cm}
	\end{subfigure}
	\caption{Streamlines and vorticity contours for the case with an aspect ratio of 3 and no eccentricity at three Reynolds numbers a) $Re=200$, b) $Re = 1000$ and c) $Re = 2000$}
    \label{fig:case2a}
\end{figure}

\begin{figure}[H]
	\centering
	\begin{subfigure}[t]{0.45\textwidth}
		\includegraphics[width=\textwidth]{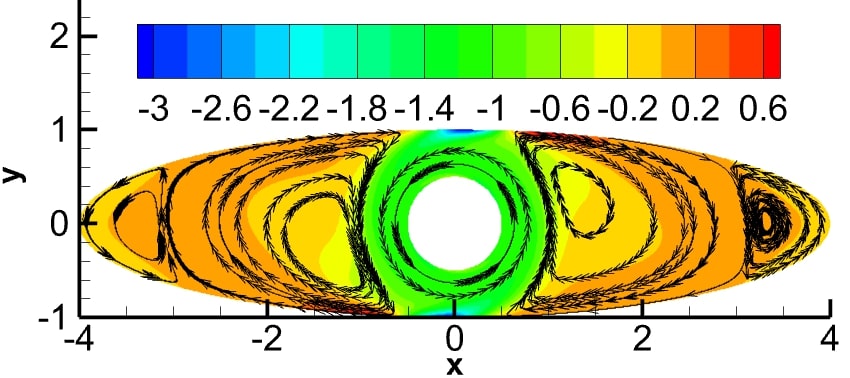}
        \caption{}
        \vspace{0.25cm}
	\end{subfigure}
	\begin{subfigure}[t]{0.45\textwidth}
		\includegraphics[width=\textwidth]{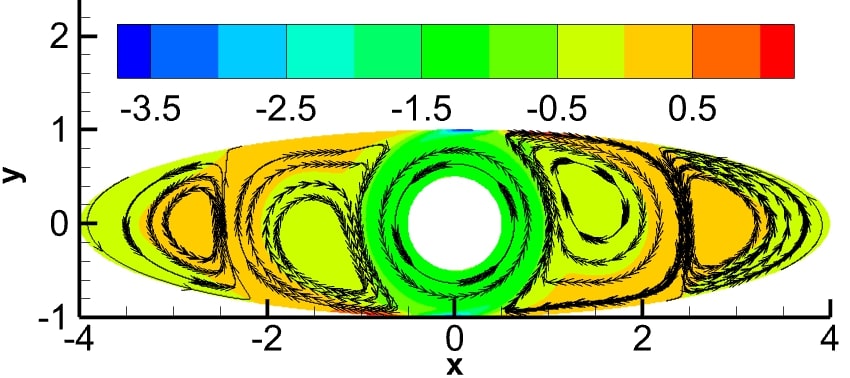}
        \caption{}
        \vspace{0.25cm}
	\end{subfigure}
	\begin{subfigure}[t]{0.45\textwidth}
		\includegraphics[width=\textwidth]{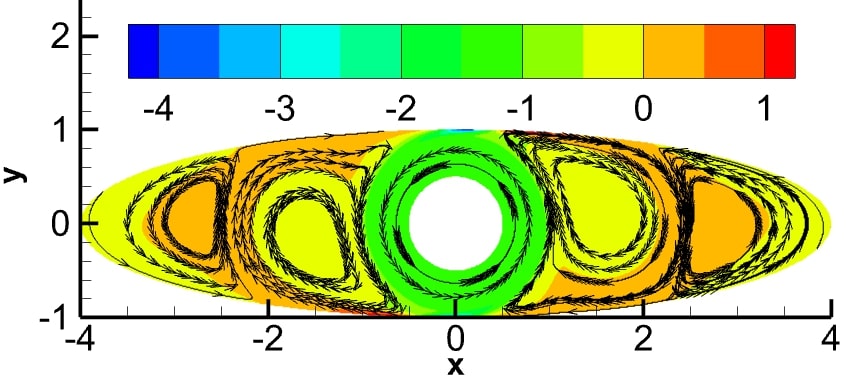}
        \caption{}
        \vspace{0.25cm}
	\end{subfigure}
	\caption{Streamlines and vorticity contours for the case with an aspect ratio of 4 and no eccentricity  at three Reynolds numbers a) $Re=200$, b) $Re = 1000$ and c) $Re = 2000$}
    \label{fig:case3a}
\end{figure}

We next consider the effects of an increase in the aspect ratio of the outer ellipse with the inner cylinder placed concentrically. Again, a number of Reynolds numbers are computed, and the antisymmetric vortices formed on both sides of the inner cylinder are studied. In addition to the base case, aspect ratios of 3 and 4 are computed with the Reynolds number ranging from 200 to 2000. As expected, the flow is antisymmetric in the direction of rotation of the inner cylinder. For higher aspect ratio, we observe the increased formation of secondary vortices. The secondary vortices appear earlier in Reynolds number compared to the base case of aspect ratio 2. As the Reynolds number is increased from 200 to 2000, the size of the primary vortex decreases, giving space for the formation of the secondary vortex (\Cref{fig:case2a}). As the aspect ratio is increased to 4 (\Cref{fig:case3a}), the secondary vortices grow as a result of the additional corner region available.


\begin{figure}[H]
    \centering
		\includegraphics[width=0.45\textwidth]{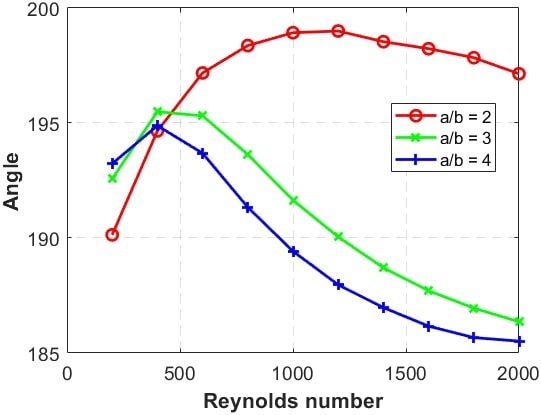}
        \caption{Primary vortex angle vs Reynolds number for no eccentricity}
        \vspace{0.25cm}
    \label{fig:angle_case123a}
\end{figure}

\Cref{fig:angle_case123a} shows the angle made by the center of the primary vortex with respect to the horizontal centerline. As seen for the case of the inner cylinder eccentricity, the increase in the aspect ratio first moves the vortex center downwards but with an increase in the Reynolds number, the vortex center moves back to be closer to the horizontal centerline. For the case of aspect ratio 2, the decrease in the angle is smaller compared to the cases of aspect ratio of 3 and 4. For aspect ratios of 3 and 4, the eye of the vortex reaches the most downward position around the Reynolds number of 500. After this value, it moves up to make an angle of approximately $5^{\text{o}}$ in horizontal.

Benchmark data for these cases are given in \Cref{tab:case 3a x,tab:case 3a y}. \Cref{tab:case 3a x} contains the vorticity and y - velocity data along the horizontal centerline and \Cref{tab:case 3a y} contains the vorticity and x - velocity data along the vertical centerline for three different Reynolds number of 400, 1000 and 2000. These accurate data can be used for assessment of future numerical studies.

\begin{table}[H]
\centering
\begin{tabular}{|c|cc|cc|cc|}
\hline
Cordinates & \multicolumn{2}{c|}{Re = 400}                      & \multicolumn{2}{c|}{Re = 1000}                     & \multicolumn{2}{c|}{Re = 2000}                     \\ \hline
x          & \multicolumn{1}{c|}{v}       & $\eta$ & \multicolumn{1}{c|}{v}       & $\eta$ & \multicolumn{1}{c|}{v}       & $\eta$ \\ \hline
-3.3350    & \multicolumn{1}{c|}{-0.0002} & 0.0006              & \multicolumn{1}{c|}{-0.0006} & 0.0002              & \multicolumn{1}{c|}{-0.0008} & 0.0001              \\ \hline
-2.6700    & \multicolumn{1}{c|}{0.0019}  & 0.0077              & \multicolumn{1}{c|}{0.0008}  & 0.0137              & \multicolumn{1}{c|}{0.0010}  & 0.0179              \\ \hline
-2.0050    & \multicolumn{1}{c|}{0.0158}  & -0.0080             & \multicolumn{1}{c|}{0.0209}  & -0.0186             & \multicolumn{1}{c|}{0.0212}  & -0.0760             \\ \hline
-1.3400    & \multicolumn{1}{c|}{-0.0067} & -0.1017             & \multicolumn{1}{c|}{-0.0133} & -0.0876             & \multicolumn{1}{c|}{-0.0154} & -0.0819             \\ \hline
-0.6750    & \multicolumn{1}{c|}{-0.5719} & -1.1134             & \multicolumn{1}{c|}{-0.5666} & -1.1492             & \multicolumn{1}{c|}{-0.5623} & -1.1752             \\ \hline
0.5000     & \multicolumn{1}{c|}{1.0000}  & -1.1394             & \multicolumn{1}{c|}{1.0000}  & -1.1270             & \multicolumn{1}{c|}{1.0000}  & -1.1317             \\ \hline
1.1650     & \multicolumn{1}{c|}{0.0306}  & -0.2062             & \multicolumn{1}{c|}{0.0263}  & -0.1049             & \multicolumn{1}{c|}{0.0249}  & -0.0580             \\ \hline
1.8300     & \multicolumn{1}{c|}{-0.0178} & -0.0393             & \multicolumn{1}{c|}{-0.0189} & -0.0819             & \multicolumn{1}{c|}{-0.0127} & -0.1013             \\ \hline
2.4950     & \multicolumn{1}{c|}{-0.0042} & 0.0105              & \multicolumn{1}{c|}{-0.0037} & 0.0219              & \multicolumn{1}{c|}{-0.0052} & 0.0320              \\ \hline
3.1600     & \multicolumn{1}{c|}{0.0001}  & 0.0016              & \multicolumn{1}{c|}{0.0008}  & 0.0016              & \multicolumn{1}{c|}{0.0012}  & 0.0020              \\ \hline
\end{tabular}%
\caption{Benchmark data for a/b = 4, y = 0.0 and e/a = 0.0, along the horizontal centerline
}
\label{tab:case 3a x}
\end{table}

\begin{table}[H]
\centering
\begin{tabular}{|c|cc|cc|cc|}
\hline
Cordinates & \multicolumn{2}{c|}{Re = 400}                      & \multicolumn{2}{c|}{Re = 1000}                     & \multicolumn{2}{c|}{Re = 2000}                     \\ \hline
y          & \multicolumn{1}{c|}{u}       & $\eta$ & \multicolumn{1}{c|}{u}       & $\eta$ & \multicolumn{1}{c|}{u}       & $\eta$ \\ \hline
-0.9750    & \multicolumn{1}{c|}{0.0726}  & -2.6688             & \multicolumn{1}{c|}{0.0766}  & -2.6913             & \multicolumn{1}{c|}{0.0787}  & -2.6013             \\ \hline
-0.8800    & \multicolumn{1}{c|}{0.2531}  & -1.1593             & \multicolumn{1}{c|}{0.2357}  & -1.0744             & \multicolumn{1}{c|}{0.2235}  & -1.1341             \\ \hline
-0.7850    & \multicolumn{1}{c|}{0.3856}  & -1.0822             & \multicolumn{1}{c|}{0.3730}  & -1.1491             & \multicolumn{1}{c|}{0.3653}  & -1.1746             \\ \hline
-0.6900    & \multicolumn{1}{c|}{0.5471}  & -1.1149             & \multicolumn{1}{c|}{0.5387}  & -1.1491             & \multicolumn{1}{c|}{0.5332}  & -1.1750             \\ \hline
-0.5950    & \multicolumn{1}{c|}{0.7465}  & -1.1190             & \multicolumn{1}{c|}{0.7413}  & -1.1491             & \multicolumn{1}{c|}{0.7381}  & -1.1769             \\ \hline
-0.5000    & \multicolumn{1}{c|}{1.0000}  & -0.6488             & \multicolumn{1}{c|}{1.0000}  & -0.8447             & \multicolumn{1}{c|}{1.0000}  & -1.0143             \\ \hline
0.5000     & \multicolumn{1}{c|}{-1.0000} & -0.6443             & \multicolumn{1}{c|}{-1.0000} & -0.8333             & \multicolumn{1}{c|}{-1.0000} & -1.0026             \\ \hline
0.5950     & \multicolumn{1}{c|}{-0.7465} & -1.1191             & \multicolumn{1}{c|}{-0.7413} & -1.1487             & \multicolumn{1}{c|}{-0.7381} & -1.1766             \\ \hline
0.6900     & \multicolumn{1}{c|}{-0.5471} & -1.1150             & \multicolumn{1}{c|}{-0.5387} & -1.1491             & \multicolumn{1}{c|}{-0.5332} & -1.1750             \\ \hline
0.7850     & \multicolumn{1}{c|}{-0.3856} & -1.0822             & \multicolumn{1}{c|}{-0.3730} & -1.1491             & \multicolumn{1}{c|}{-0.3653} & -1.1746             \\ \hline
0.8800     & \multicolumn{1}{c|}{-0.2531} & -1.1594             & \multicolumn{1}{c|}{-0.2357} & -1.0742             & \multicolumn{1}{c|}{-0.2235} & -1.1338             \\ \hline
0.9750     & \multicolumn{1}{c|}{-0.0726} & -2.6689             & \multicolumn{1}{c|}{-0.0766} & -2.6914             & \multicolumn{1}{c|}{-0.0787} & -2.6019             \\ \hline
\end{tabular}%
\caption{Benchmark data for a/b = 4, x = 0.0 and e/a = 0.0, along the vertical centerline}
\label{tab:case 3a y}
\end{table}

\section{EFFECTS OF COMBINED ECCENTRICITY AND ASPECT RATIO}
\label{sec:effect_of_e_and_ar}
\begin{figure}[H]
	\centering
	\begin{subfigure}[t]{0.45\textwidth}
		\includegraphics[width=\textwidth]{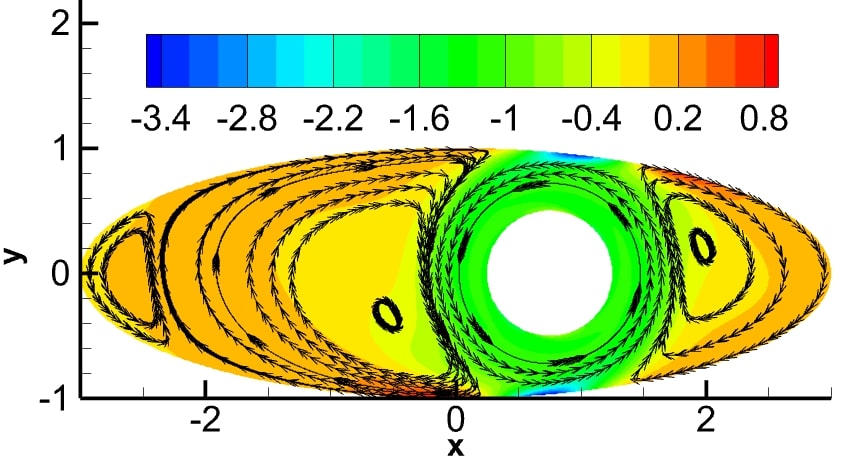}
        \caption{}
        \vspace{0.25cm}
	\end{subfigure}
	\begin{subfigure}[t]{0.45\textwidth}
		\includegraphics[width=\textwidth]{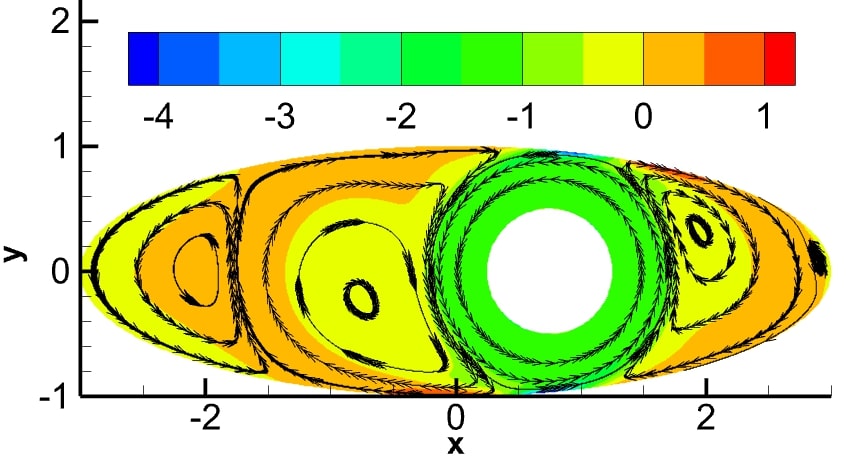}
        \caption{}
        \vspace{0.25cm}
	\end{subfigure}
	\begin{subfigure}[t]{0.45\textwidth}
		\includegraphics[width=\textwidth]{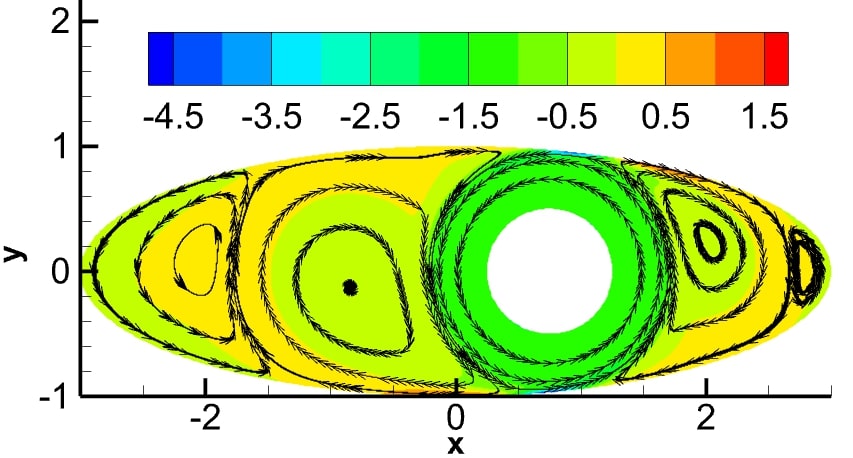}
        \caption{}
        \vspace{0.25cm}
	\end{subfigure}
	\caption{Streamlines and vorticity contours for the case with an aspect ratio of 3 and an eccentricity ratio of 0.25 at three Reynolds numbers a) $Re=200$, b) $Re = 1000$ and c) $Re = 2000$}
    \label{fig:case2b}
\end{figure}

\begin{figure}[H]
	\centering
	\begin{subfigure}[t]{0.45\textwidth}
		\includegraphics[width=\textwidth]{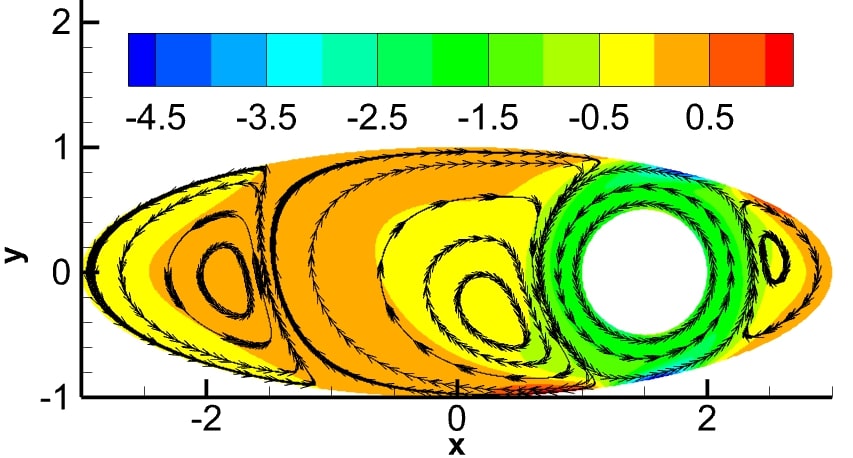}
        \caption{}
        \vspace{0.25cm}
	\end{subfigure}
	\begin{subfigure}[t]{0.45\textwidth}
		\includegraphics[width=\textwidth]{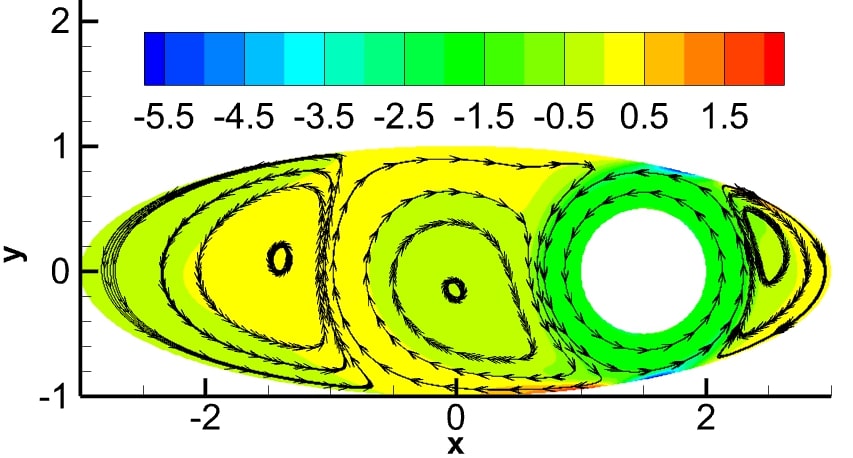}
        \caption{}
        \vspace{0.25cm}
	\end{subfigure}
	\begin{subfigure}[t]{0.45\textwidth}
		\includegraphics[width=\textwidth]{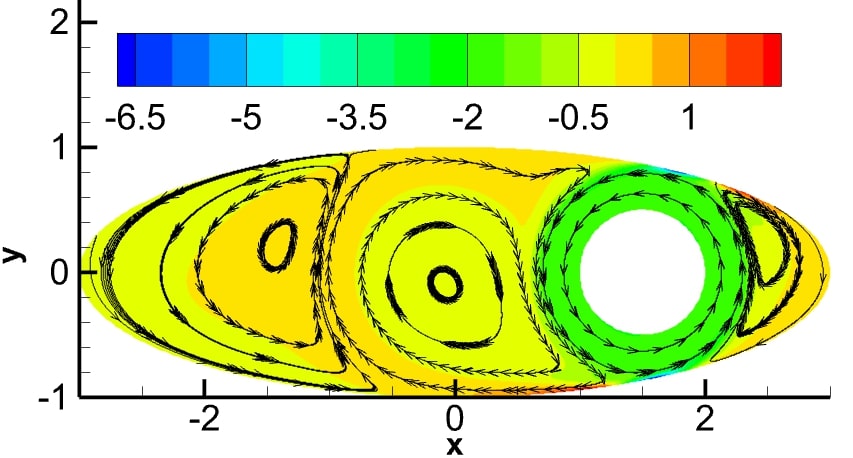}
        \caption{}
        \vspace{0.25cm}
	\end{subfigure}
	\caption{Streamlines and vorticity contours for the case with an aspect ratio of 3 and an eccentricity ratio of 0.5 at three Reynolds numbers a) $Re=200$, b) $Re = 1000$ and c) $Re = 2000$}
    \label{fig:case2c}
\end{figure}

\begin{figure}[H]
	\centering
	\begin{subfigure}[t]{0.45\textwidth}
		\includegraphics[width=\textwidth]{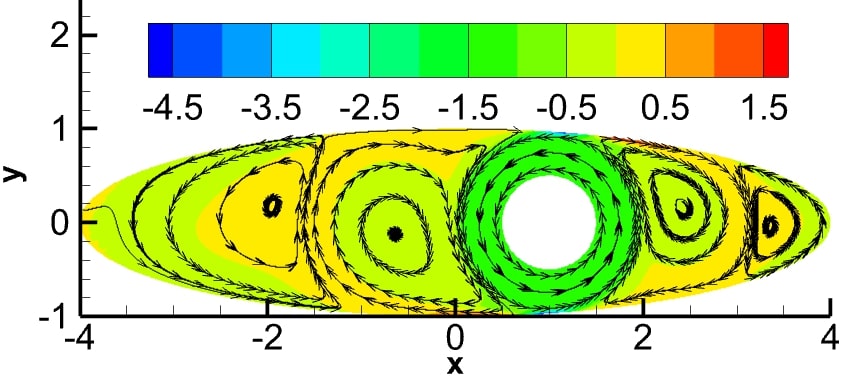}
        \caption{}
        \vspace{0.25cm}
	\end{subfigure}
	\begin{subfigure}[t]{0.45\textwidth}
		\includegraphics[width=\textwidth]{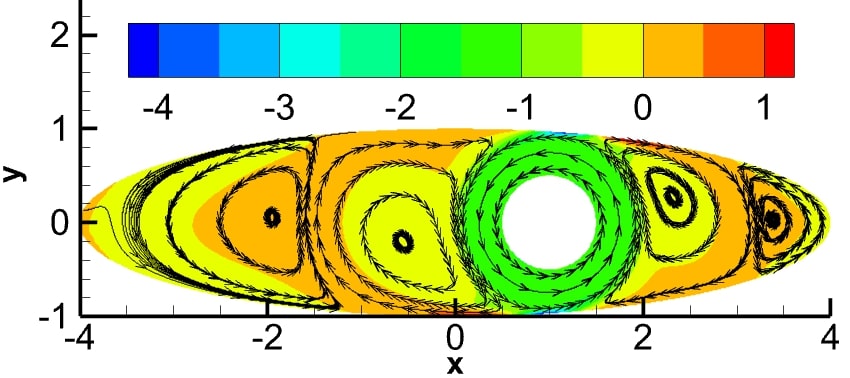}
        \caption{}
        \vspace{0.25cm}
	\end{subfigure}
	\begin{subfigure}[t]{0.45\textwidth}
		\includegraphics[width=\textwidth]{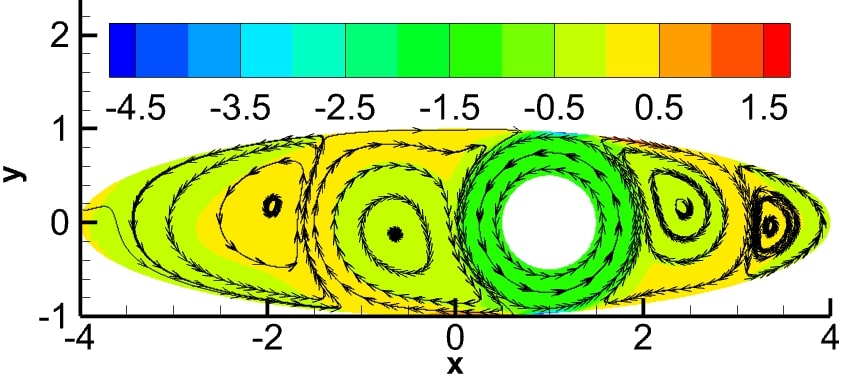}
        \caption{}
        \vspace{0.25cm}
	\end{subfigure}
	\caption{Streamlines and vorticity contours for the case with an aspect ratio of 4 and an eccentricity ratio of 0.25 at three Reynolds numbers a) $Re=200$, b) $Re = 1000$ and c) $Re = 2000$}
    \label{fig:case3b}
\end{figure}

\begin{figure}[H]
	\centering
	\begin{subfigure}[t]{0.45\textwidth}
		\includegraphics[width=\textwidth]{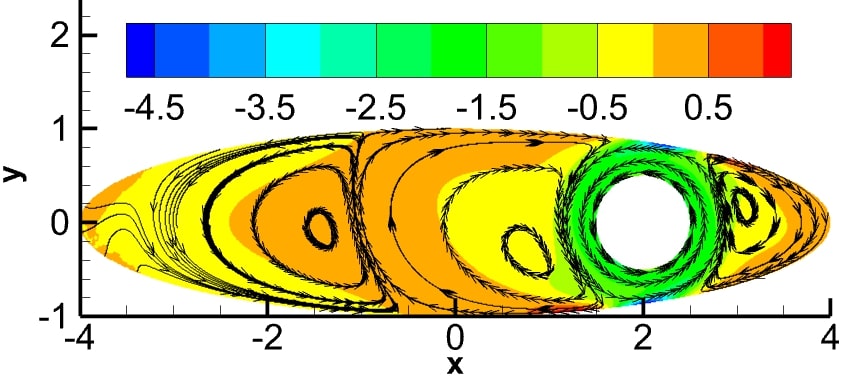}
        \caption{}
        \vspace{0.25cm}
	\end{subfigure}
	\begin{subfigure}[t]{0.45\textwidth}
		\includegraphics[width=\textwidth]{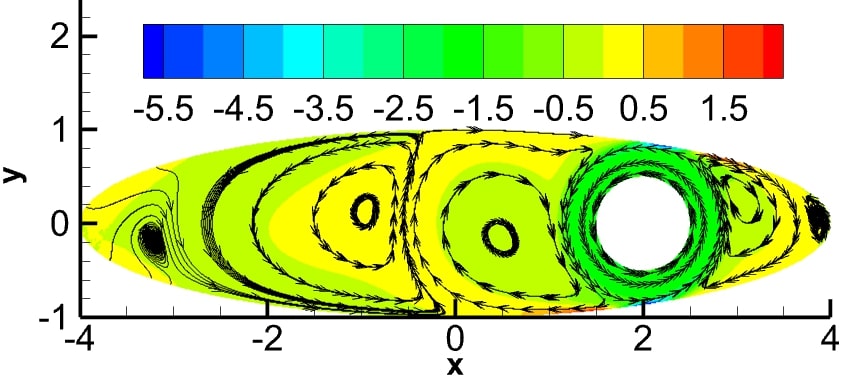}
        \caption{}
        \vspace{0.25cm}
	\end{subfigure}
	\begin{subfigure}[t]{0.45\textwidth}
		\includegraphics[width=\textwidth]{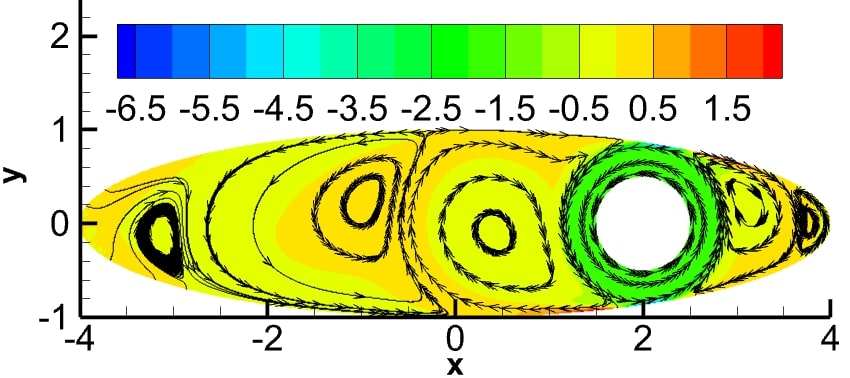}
        \caption{}
        \vspace{0.25cm}
	\end{subfigure}
	\caption{Streamlines and vorticity contours for the case with an aspect ratio of 4 and an eccentricity ratio of 0.5 at three Reynolds numbers a) $Re=200$, b) $Re = 1000$ and c) $Re = 2000$}
    \label{fig:case3c}
\end{figure}

When both eccentricity and aspect ratio are increased, secondary vortices begin to appear in the wide gap region even at low Reynolds numbers (\Cref{fig:case2b,fig:case2c,fig:case3b,fig:case3c}). In the small gap region for an aspect ratio of 3 we see secondary vortices for an eccentricity ratio of 0.25, while they are not present for the cases where the eccentricity ratio is 0.5. For the higher aspect ratio of 4, and for the case of eccentricity ratio 0.25, the primary and secondary vortices exist for all the Reynolds numbers for which the simulations are run. For an eccentricity ratio of 0.5, the secondary vortices appear in the small gap region as the Reynolds number is increased slowly. In the wide gap region a tertiary vortex appears as the Reynolds number is increased.

\begin{figure}[H]
	\centering
	\begin{subfigure}[t]{0.45\textwidth}
		\includegraphics[width=\textwidth]{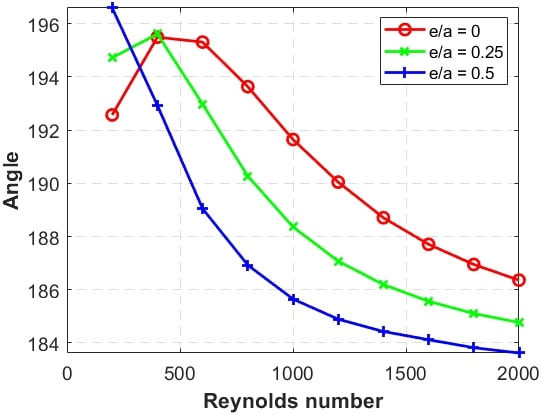}
        \caption{Case 2: $a/b = 3$}
	\end{subfigure}
	\hspace{0.05\textwidth}
	\begin{subfigure}[t]{0.45\textwidth}
		\includegraphics[width=\textwidth]{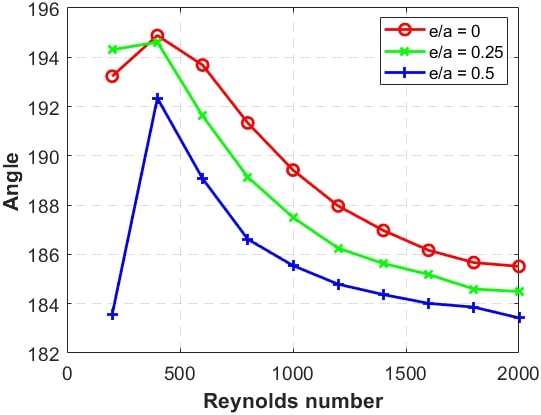}
        \caption{Case 3: $a/b = 4$} \vspace{0.25cm}
	\end{subfigure}
	\caption{Primary vortex eye angle vs Reynolds number for eccentricity ratios of 0, 0.25 and 0.5 and aspect ratios: a) $a/b = 3$; b) $a/b = 4$}
    \label{fig:angle_case23}
\end{figure}

Angles of eyes of primary vortices in the cases of higher aspect ratio and eccentricity follow previously observed trends as the Reynolds number is increased. The  maximum angle is seen to be at around Reynolds number of 400 for most of the cases (\Cref{fig:angle_case23}).

Benchmark data for the case of an aspect ratio of 4 and an eccentricity ratio of 0.5, i.e. the extreme case simulated are given in \Cref{tab:case 3c x,tab:case 3c y}. The data given are vorticity and y velocity along the horizontal centerline for the cases of Reynolds number 400, 1000 and 2000 in \Cref{tab:case 3c x} and vorticity and x velocity along the vertical centerline for the same Reynolds numbers are in \Cref{tab:case 3c y}.

\begin{table}[H]
\centering
\begin{tabular}{|c|cc|cc|cc|}
\hline
Cordinates & \multicolumn{2}{c|}{Re = 400}                      & \multicolumn{2}{c|}{Re = 1000}                     & \multicolumn{2}{c|}{Re = 2000}                     \\ \hline
x          & \multicolumn{1}{c|}{v}       & $\eta$ & \multicolumn{1}{c|}{v}       & $\eta$ & \multicolumn{1}{c|}{v}       & $\eta$ \\ \hline
-2.9550    & \multicolumn{1}{c|}{0.0000}  & -0.0001             & \multicolumn{1}{c|}{0.0000}  & -0.0001             & \multicolumn{1}{c|}{0.0000}  & -0.0002             \\ \hline
-1.9100    & \multicolumn{1}{c|}{-0.0005} & 0.0003              & \multicolumn{1}{c|}{-0.0009} & 0.0002              & \multicolumn{1}{c|}{-0.0010} & -0.0001             \\ \hline
-0.8650    & \multicolumn{1}{c|}{0.0014}  & 0.0117              & \multicolumn{1}{c|}{0.0011}  & 0.0175              & \multicolumn{1}{c|}{0.0016}  & 0.0164              \\ \hline
0.1800     & \multicolumn{1}{c|}{0.0271}  & -0.0776             & \multicolumn{1}{c|}{0.0196}  & -0.1340             & \multicolumn{1}{c|}{0.0112}  & -0.0980             \\ \hline
1.2250     & \multicolumn{1}{c|}{-0.2642} & -1.5388             & \multicolumn{1}{c|}{-0.2473} & -1.7264             & \multicolumn{1}{c|}{-0.2375} & -1.8285             \\ \hline
2.5000     & \multicolumn{1}{c|}{1.0000}  & -1.8087             & \multicolumn{1}{c|}{1.0000}  & -1.7872             & \multicolumn{1}{c|}{1.0000}  & -1.7793             \\ \hline
2.7850     & \multicolumn{1}{c|}{0.2395}  & -1.5801             & \multicolumn{1}{c|}{0.2247}  & -1.7536             & \multicolumn{1}{c|}{0.2158}  & -1.8452             \\ \hline
3.0700     & \multicolumn{1}{c|}{0.0051}  & -0.2307             & \multicolumn{1}{c|}{0.0059}  & -0.1518             & \multicolumn{1}{c|}{0.0119}  & -0.1682             \\ \hline
3.3550     & \multicolumn{1}{c|}{-0.0213} & -0.0561             & \multicolumn{1}{c|}{-0.0197} & -0.0379             & \multicolumn{1}{c|}{-0.0226} & -0.0949             \\ \hline
3.6400     & \multicolumn{1}{c|}{-0.0090} & 0.0232              & \multicolumn{1}{c|}{-0.0054} & 0.0357              & \multicolumn{1}{c|}{-0.0052} & 0.0584              \\ \hline
3.9250     & \multicolumn{1}{c|}{-0.0005} & 0.0091              & \multicolumn{1}{c|}{0.0002}  & 0.0011              & \multicolumn{1}{c|}{0.0007}  & -0.0038             \\ \hline
\end{tabular}%
\caption{Benchmark data for a/b = 4, y = 0.0, e/a = 0.5, along horizontal centerline}
\label{tab:case 3c x}
\end{table}

\begin{table}[H]
\centering
\begin{tabular}{|c|cc|cc|cc|}
\hline
Cordinates & \multicolumn{2}{c|}{Re = 400}                      & \multicolumn{2}{c|}{Re = 1000}                     & \multicolumn{2}{c|}{Re = 2000}                     \\ \hline
y          & \multicolumn{1}{c|}{u}       & $\eta$ & \multicolumn{1}{c|}{u}       & $\eta$ & \multicolumn{1}{c|}{u}       & $\eta$ \\ \hline
-0.8477    & \multicolumn{1}{c|}{0.0851}  & -4.1390             & \multicolumn{1}{c|}{-0.0069} & 0.0912              & \multicolumn{1}{c|}{0.0889}  & -3.5370             \\ \hline
-0.7782    & \multicolumn{1}{c|}{0.2718}  & -1.6167             & \multicolumn{1}{c|}{-0.0129} & 0.1130              & \multicolumn{1}{c|}{0.2331}  & -1.6641             \\ \hline
-0.7086    & \multicolumn{1}{c|}{0.4081}  & -1.5860             & \multicolumn{1}{c|}{-0.0188} & 0.1129              & \multicolumn{1}{c|}{0.3838}  & -1.8031             \\ \hline
-0.6391    & \multicolumn{1}{c|}{0.5716}  & -1.6857             & \multicolumn{1}{c|}{-0.0233} & 0.0894              & \multicolumn{1}{c|}{0.5575}  & -1.8017             \\ \hline
-0.5696    & \multicolumn{1}{c|}{0.7660}  & -1.7345             & \multicolumn{1}{c|}{-0.0255} & 0.0489              & \multicolumn{1}{c|}{0.7585}  & -1.8069             \\ \hline
-0.5000    & \multicolumn{1}{c|}{1.0000}  & -1.0881             & \multicolumn{1}{c|}{-0.0251} & 0.0032              & \multicolumn{1}{c|}{1.0000}  & -1.7289             \\ \hline
0.5000     & \multicolumn{1}{c|}{-1.0000} & -0.9271             & \multicolumn{1}{c|}{0.0180}  & 0.0006              & \multicolumn{1}{c|}{-1.0000} & -1.3666             \\ \hline
0.5696     & \multicolumn{1}{c|}{-0.7780} & -1.7367             & \multicolumn{1}{c|}{0.0171}  & 0.0104              & \multicolumn{1}{c|}{-0.7627} & -1.8031             \\ \hline
0.6391     & \multicolumn{1}{c|}{-0.5833} & -1.6931             & \multicolumn{1}{c|}{0.0157}  & 0.0174              & \multicolumn{1}{c|}{-0.5621} & -1.8015             \\ \hline
0.7086     & \multicolumn{1}{c|}{-0.4194} & -1.6631             & \multicolumn{1}{c|}{0.0139}  & 0.0223              & \multicolumn{1}{c|}{-0.3895} & -1.7990             \\ \hline
0.7782     & \multicolumn{1}{c|}{-0.2609} & -2.3510             & \multicolumn{1}{c|}{0.0120}  & 0.0275              & \multicolumn{1}{c|}{-0.2392} & -1.7624             \\ \hline
0.8477     & \multicolumn{1}{c|}{-0.0531} & -3.0220             & \multicolumn{1}{c|}{0.0096}  & 0.0374              & \multicolumn{1}{c|}{-0.0510} & -3.1516             \\ \hline
\end{tabular}%
\caption{Benchmark data for a/b = 4, x = 2.0, e/a = 0.5, along vertical centerline}
\label{tab:case 3c y}
\end{table}

The torque required to rotate the inner cylinder can be computed from the shear stresses exerted by the fluid. This is further non-dimensionalized as discussed by \citet{fasel_booz_1984}.
\begin{equation}
    T_n = \frac{T Re}{\omega_i R_i^2 L}
    \label{eqn:non dimensional torque}
\end{equation}

\begin{figure}[H]
	\centering
	\begin{subfigure}[t]{0.45\textwidth}
		\includegraphics[width=\textwidth]{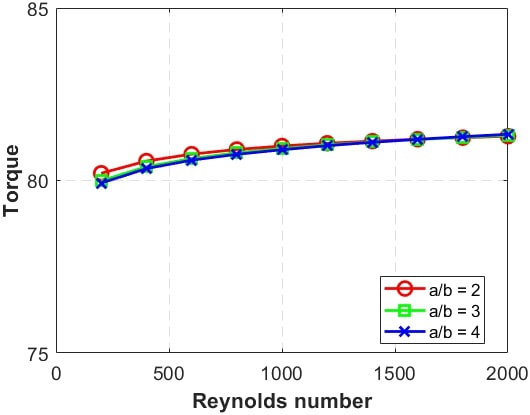}
         \caption{$e/a = 0$} \vspace{0.25cm}
	\end{subfigure}
	\begin{subfigure}[t]{0.45\textwidth}
		\includegraphics[width=\textwidth]{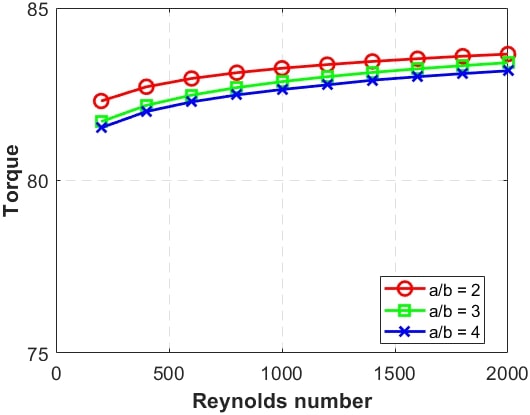}
         \caption{$e/a = 0.25$} \vspace{0.25cm}
	\end{subfigure}
	\hspace{0.05\textwidth}
	\begin{subfigure}[t]{0.45\textwidth}
		\includegraphics[width=\textwidth]{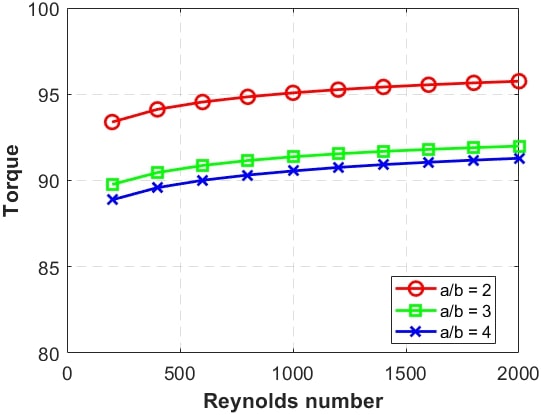}
        \caption{$e/a = 0.5$}
	\end{subfigure}

	\caption{Torque vs Reynolds number plots for all the test cases. (a) For $e/a = 0$, three $a/b$ values of (2, 3 and 4); (b) For $e/a = 0.25$, three $a/b$ values of (2, 3 and 4); and (c) For $e/a = 0.50$, three $a/b$ values of (2, 3 and 4); }
    \label{fig:Re vs Torque abc}
\end{figure}

The magnitude of torque is influenced by the eccentricity and aspect ratio since the flow fields and the viscous shear will be different for each geometry.
\Cref{fig:Re vs Torque abc} shows the variation of torque with Reynolds number for the three aspect ratios with no eccentricity. It is seen that the torque is nearly independent of aspect ratio, and slightly increases with Reynolds number. For eccentricity ratios of 0.25 and 0.5, we see that the torque decreases for aspect ratios of 3 and 4 by as much as 5\% in the case of e/a = 0.5.

\begin{figure}[H]
	\centering
	\begin{subfigure}[t]{0.45\textwidth}
		\includegraphics[width=\textwidth]{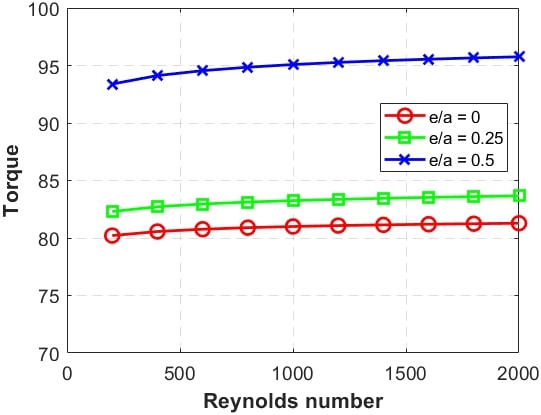}
        \caption{Case 1: $a/b = 2$}
	\end{subfigure}
	\hspace{0.05\textwidth}
	\begin{subfigure}[t]{0.45\textwidth}
		\includegraphics[width=\textwidth]{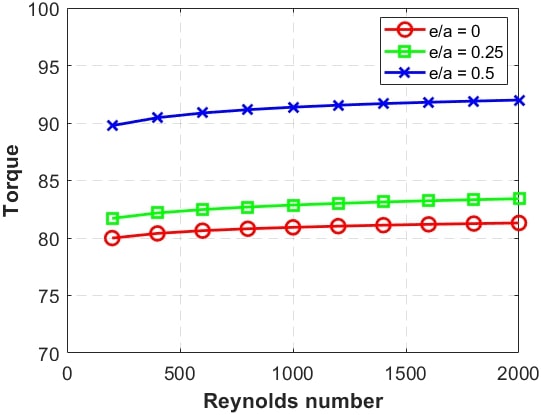}
        \caption{Case 2: $a/b = 3$} \vspace{0.25cm}
	\end{subfigure}
	\begin{subfigure}[t]{0.45\textwidth}
		\includegraphics[width=\textwidth]{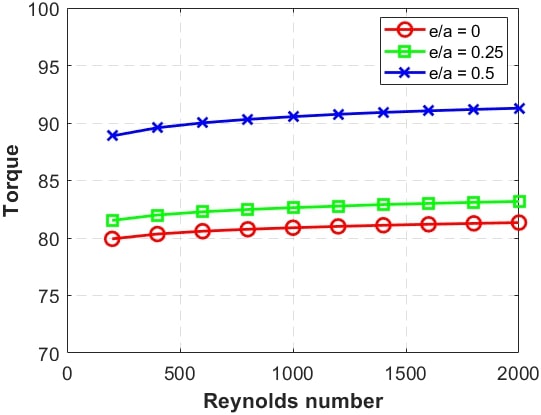}
        \caption{Case 3: $a/b = 4$}
	\end{subfigure}
	\hspace{0.05\textwidth}
	\caption{Torque vs Reynolds number plots for three eccentricity ratios of 0, 0.25 and 0.5 for aspect ratios of (a) $a/b = 2$; (b) $a/b = 3$; (c) $a/b = 4$; }
    \label{fig:Re vs Torque 123}
\end{figure}

\Cref{fig:Re vs Torque 123} plots the torque vs Reynolds number for different aspect ratios with results for the three eccentricities plotted together. In this way, we focus on the effects of eccentricity. We observe that the case of e/a = 0.5 requires the largest torque of all the three cases. The torque is nearly constant with Reynolds number.  The effect of eccentricity is more pronounced in the case of lower aspect ratios. The higher the aspect ratio, the magnitude of torque appears to approach a constant value with increasing Reynolds numbers.

\section{SUMMARY}
\label{sec:summary}
In this paper, we have analyzed the fluid flow between a rotating circular cylinders inside the fixed elliptical enclosure. A novel high accuracy meshless method is used to discretize the governing equations. The partial derivatives in the Navier-Stokes equations are evaluated by interpolating the velocities and pressure by Polyharmonic splines with appended polynomials. A semi-implicit method is used to solve the time-dependent equations to the steady state. Computations have been performed for a number of parameters, varying the Reynolds number, aspect ratio of the outer ellipse and the placement of the inner cylinder concentrically or eccentrically. For each case, grid-independent velocities and pressures are computed at the steady state. The observed flow patterns and vorticity variations are systematically studied.   The velocities and vorticity at selected lines for a few cases are provided in the tabular form to serve as benchmark data for future numerical studies.

The rotation of the inner cylinder generates symmetrical (for concentric placement of the inner cylinder) and unsymmetric (for eccentric placement) primary vortices on either side of the major axis. In the direction of the minor axis (vertical direction), the tangential rotational velocity is seen in the narrow gaps. The primary vortices in the larger gaps are similar to those in shear driven cavity flows studied in complex enclosures. These in turn are seen to drive weaker and smaller secondary vortices between the primary vortices and the wall of the ellipse. When the inner cylinder is placed eccentrically in the ellipse, the flow becomes unsymmetric with different sizes of the primary vortices as expected. For larger eccentricity, the primary vortex on the small gap size shrinks but a secondary vortex appears on the side of the larger gap. As the aspect ratio of the outer ellipse is increased to 3 and 4, the secondary vortices increase in size and strength, especially at higher Reynolds numbers. This is similar to the formation of multiple vortices in triangular cavities as the apex angle of the triangle is decreased. As the aspect ratio is increased, the secondary vortices are more clearly seen at lower Reynolds numbers. Small tertiary vortices are also observed for the aspect ratio of 4 and Reynolds number of 2000 with the inner cylinder placed eccentrically. These vortices are quite weak as in a triangular cavity-driven flow. We have documented the vorticity contours and angles made by the vortex centers for each of the cases computed.

The torque required to keep the inner cylinder rotating at constant velocity is found to be dependent on the eccentricity ratio and not much on the aspect ratio of the elliptical enclosure. It is also found that the non-dimensionalized torque is a weak function of the Reynolds number.

\section*{Data Availability}
The data that supports the findings of this study are available within the article.

\section*{Author Declarations}
The authors have no conflicts to disclose.

\bibliography{References}

\end{document}